\documentclass[letterpaper, reprint, aip, pop, superscriptaddress, longbibliography, 
    noshowpacs, floatfix, noshowkeys]{revtex4-1}

\usepackage{amsmath}
\usepackage{graphicx}
\usepackage[pass]{geometry}

\newcommand\dif{\ensuremath{\mathrm{d}}}

\newcommand\xn{\ensuremath{\mathbf{x}_n}}
\newcommand\dotxn{\ensuremath{\dot{\mathbf{x}}_n}}
\newcommand\U{\ensuremath{\mathbf{U}}}
\newcommand\V{\ensuremath{\mathbf{V}}}
\newcommand\Vi{\ensuremath{\mathbf{V}_i}}
\newcommand\Ve{\ensuremath{\mathbf{V}_e}}
\newcommand\J{\ensuremath{\mathbf{J}}}

\newcommand\B{\ensuremath{\mathbf{B}}}
\newcommand\E{\ensuremath{\mathbf{E}}}

\newcommand\x{\ensuremath{\mathbf{x}}}

\renewcommand\r{\ensuremath{\delta\mathbf{x}}}
\newcommand\dbdt{\ensuremath{\frac{\partial \B}{\partial t}}}
\newcommand\dbdtxnA{\ensuremath{\left.\frac{\partial \B}{\partial t}\right|_{\xn(t)}}}
\newcommand\dbdtxnB{\ensuremath{\frac{\partial\B_n}{\partial t}}}
\newcommand\dbdtxn{\dbdtxnB}

\newcommand\zhat{\ensuremath{\hat{\mathbf{z}}}}
\newcommand\M{\ensuremath{\boldsymbol{\mathsf{M}}}}
\newcommand\Minv{\ensuremath{\M^{-1}}}
\newcommand{\sgn}{\mathop{\mathrm{sgn}}}
\newcommand{\rank}{\mathop{\mathrm{rank}}}
\newcommand{\nullity}{\mathop{\mathrm{nullity}}}

\newcommand{\dbx}{\ensuremath{\dot{B}_x}}
\newcommand{\dby}{\ensuremath{\dot{B}_y}}
\newcommand{\dbz}{\ensuremath{\dot{B}_z}}

\renewcommand\H{\ensuremath{\boldsymbol{\mathsf{H}}}}
\newcommand\Hx{\ensuremath{\boldsymbol{\mathsf{H}}_x}}
\newcommand\Hy{\ensuremath{\boldsymbol{\mathsf{H}}_y}}
\newcommand\Hz{\ensuremath{\boldsymbol{\mathsf{H}}_z}}

\newcommand{\Aorminus}{\ensuremath{-}}
\newcommand{\Borplus}{\ensuremath{+}}

\newcommand{\gammaA}{\ensuremath{\gamma_{\Aorminus}}}
\newcommand{\gammaB}{\ensuremath{\gamma_{\Borplus}}}
\newcommand{\SigmaA}{\ensuremath{\Sigma_{\Aorminus}}}
\newcommand{\SigmaB}{\ensuremath{\Sigma_{\Borplus}}}

\begin{document}

\title{The appearance, motion, and disappearance of three-dimensional magnetic null points}
\author{Nicholas A. Murphy}
\email[]{namurphy@cfa.harvard.edu}
\affiliation{Harvard-Smithsonian Center for Astrophysics, Cambridge, Massachusetts 02138, USA}

\author{Clare E. Parnell}
\affiliation{School of Mathematics and Statistics, University of St Andrews, North Haugh, St Andrews, Fife, KY16 9SS, UK}

\author{Andrew L. Haynes}
\affiliation{School of Mathematics and Statistics, University of St Andrews, North Haugh, St Andrews, Fife, KY16 9SS, UK}

\date{\today}


\begin{abstract}

While theoretical models and simulations of magnetic reconnection often assume symmetry such that the magnetic null point when present is co-located with a flow stagnation point, the introduction of asymmetry typically leads to non-ideal flows across the null point. To understand this behavior, we present exact expressions for the motion of three-dimensional linear null points.  The most general expression shows that linear null points move in the direction along which the vector field and its time derivative are antiparallel.  Null point motion in resistive magnetohydrodynamics results from advection by the bulk plasma flow and resistive diffusion of the magnetic field, which allows non-ideal flows across topological boundaries.  Null point motion is described intrinsically by parameters evaluated locally; however, global dynamics help set the local conditions at the null point.  During a bifurcation of a degenerate null point into a null-null pair or the reverse, the instantaneous velocity of separation or convergence of the null-null pair will typically be infinite along the null space of the Jacobian matrix of the magnetic field, but with finite components in the directions orthogonal to the null space.  Not all bifurcating null-null pairs are connected by a separator.  Furthermore, except under special circumstances, there will not exist a straight line separator connecting a bifurcating null-null pair.  The motion of separators cannot be described using solely local parameters, because the identification of a particular field line as a separator may change as a result of non-ideal behavior elsewhere along the field line.

\end{abstract}


\maketitle 

\section{Introduction}

Magnetic reconnection\cite{PF00, *birn:2007:book, *zweibel:2009, *yamada:2010} frequently occurs at and around magnetic null points: locations where the magnetic field strength equals zero.\cite{cowley:1973, fukao:1975, greene:1988, lau:1990, parnell:1996}  Magnetospheric null points have been identified using multipoint \emph{in situ} measurements as the nulls pass through the spacecraft constellation.\cite{xiao:2006, *xiao:2007, *he:2008, *wendel:2013, *rguo:2013, *fu:2015, *olshevsky:2015}  Null points in the solar atmosphere have been identified through extrapolation of the photospheric magnetic field and morphology in coronal emission.\cite{filippov:1999, *aulanier:2000, *zhao:2008, *longcope:2009, *freed:2015, *edwards:2015, *masson:2009, *demoulin:1994, *barnes:2007, *Close:2004, *Regnier:2008}  Numerical simulations of magnetic reconnection and plasma turbulence at low guide fields frequently show the formation and evolution of null points,\cite{servidio:2009, *servidio:2010} as do numerical experiments of typical solar events such as flux emergence.\cite{Maclean:2009, *Parnell:2010B}

Two-dimensional, non-degenerate magnetic null points are classified as X-type or O-type depending on the local magnetic field structure.  If we define \M\ as the Jacobian matrix of the magnetic field at the null point, then a null point will be X-type if $\det\M<0$, O-type if $\det\M>0$, and degenerate if $\det\M=0$\@.  Magnetic reconnection in two dimensions can only occur at null points.\cite[e.g.,][]{priest:2003, pontin:2011:review}  

In three dimensions, the structure of non-degenerate magnetic null points is significantly more complex.\cite{cowley:1973, fukao:1975, greene:1988, lau:1990, parnell:1996}  Null lines and null planes are structurally unstable and unlikely to exist in real systems.\cite[e.g.,][]{greene:1988, hornig:1996}  The magnetic field structure around a linear three-dimensional null point includes separatrix surfaces (or fans) of infinitely many field lines that originate (or terminate) at the null, and two spine field lines that end (or begin) at the null.  A negative (or type A) null point has separatrix surface field lines heading inward toward the null point with spine field lines heading outward from the null point.  In contrast, a positive (or type B) null point has separatrix surface field lines heading outward away from the null point and spine field lines heading inward toward the null point. 

Separators (also known as X-lines by some in the magnetospheric community) are magnetic field lines that connect two nulls. Separators that include a spine field line are not structurally stable, so separators in real systems will almost always be given by the intersection of two separatrix surfaces.  Null points, separatrix surfaces, spines, and separators are the topological boundaries that divide the magnetic field into distinct domains and are therefore preferred locations for magnetic reconnection.\cite{longcope:2005, haynes:2010, Parnell:2010A, Parnell:2010B}  Three-dimensional magnetic reconnection can also occur without nulls,\cite{Hesse:1988, Schindler:1988, priest:1992B, *aulanier:2006, *janvier:2013:III, pontin:2011:review} especially in regions such as quasi-separatrix layers where the magnetic connectivity changes quickly.

Motion of magnetic null points and reconnection regions occurs during any realistic occurrence of magnetic reconnection.  In Earth's magnetosphere, X-line retreat has been observed in the magnetotail\cite{forbes:1981, *hasegawa:2008, *oka:2011, *xcao:2012} and poleward of the cusp.\cite{wilder:2014}  At the dayside magnetopause \cite{swisdak:2003, *phan:2013} and in tokamaks,\cite{rogers:1995, *beidler:2011} the combination of a plasma-pressure gradient and a guide field leads to diamagnetic drifting of the reconnection site that can suppress reconnection.  Laboratory experiments frequently show reconnection site motion and asymmetry, often due to geometry or the Hall effect.\cite{inomoto:counter, *yoo:2014, murphy:mrx, lukin:2011}  During solar flares, the reconnection site often rises with time as the flare loops grow and can also show transverse motions.\cite[e.g.,][]{forbes:1996, *savage:2010}

Theoretical models of magnetic reconnection often assume symmetry such that each magnetic null coincides with a flow stagnation point in the reference frame of the system.  When asymmetry is introduced, there is in general a separation between these two points,\cite{cassak:asym, *cassak:hall, *cassak:dissipation, murphy:mrx, murphy:asym, murphy:retreat, oka:2008, murphy:partialasym} and in some cases a stagnation point might not even exist near a null point.\cite{murphy:double} In all of these situations, there will generally be plasma flow across the magnetic null and the null will change position. Interestingly, the velocity of a null point will generally not equal the plasma flow velocity at the null point.\cite{oka:2008, murphy:retreat, murphy:partialasym, murphy:double}  This effect is similar to the flow-through mode of reconnection.\cite{siscoe:2002, *maynard:2012}  During asymmetric magnetic reconnection in partially ionized plasmas, there may exist neutral flow through the current sheet from the weak magnetic field (high neutral pressure) side to the strong magnetic field (low neutral pressure) side due to the neutral pressure gradient.\cite{murphy:partialasym}

In previous work,\cite{murphy:retreat} we derived an exact expression for the motion of an X-line when its location is constrained to one dimension by symmetry.  In resistive magnetohydrodynamics (MHD), X-line motion results from a combination of advection by the bulk plasma flow and resistive diffusion of the normal component of the magnetic field.  In this work, we present exact expressions for the motion of linear null points in three dimensions and discuss the typical properties of the bifurcations of degenerate magnetic null points.  Section \ref{linear} contains a derivation of the motion of linear null points in a vector field.  Section \ref{magnetic} uses the results from Section \ref{linear} to describe the motion of magnetic null points.  Section \ref{bifurcation} considers the local bifurcation properties of magnetic null points and provides 
three examples.  Section \ref{discussion} contains a summary and discussion of this work.

\section{Motion of Linear Null Points in an Arbitrary Vector Field\label{linear}}

We define $\xn(t)$ as the time-dependent position of an isolated null point in a vector field $\B(\mathbf{x},t)$\@. We define $\mathbf{B}_n(\xn(t),t)$ as the value of the vector field at the null; while $\B_n\equiv 0$ for all time, $\dbdtxnA \equiv \dbdtxnB \neq 0$ when the null point is moving.
We define \U\ to be the velocity of this null,
\begin{equation}
    \U \equiv \frac{\dif \xn}{\dif t}. \label{udef}
\end{equation}
The local structure of a non-degenerate null point can be found by taking a Taylor expansion and keeping the linear terms.\cite{cowley:1973, fukao:1975, greene:1988, lau:1990, parnell:1996}  The linear structure is then given by
\begin{equation}
\B = \M\cdot\r,
\label{BMr}
\end{equation}
where $\r\equiv\x-\xn$.  The elements of the Jacobian matrix \M\ evaluated at the null are given by
\begin{equation}
M_{ij} = \partial_j B_i, \label{jacobiandef}
\end{equation}
where $i$ is the row index and $j$ is the column index. The trace of \M\ equals zero when $\nabla\cdot\B=0$, and $\M=(\nabla\B)^T$.

Next we take the derivative following the motion of the null,
\begin{equation}
  \dbdtxn + \left. \U \cdot \nabla \B \right|_{\xn} = 0. \label{convectiveder}
\end{equation}
This expression gives the total derivative of the magnetic field at the null point using the null's velocity in an arbitrary reference frame.  This derivative equals zero because the magnetic field at the null by definition does not deviate from zero as we are following it.  By solving for \U\ in Eq.\ \ref{convectiveder}, we arrive at the most general expression for the velocity of the null point\cite{ [][{. A similar expression for null point motion is given by Eq.\ 15 of this reference; however, this expression contains a sign error.}]greene:1993, [][{. See Eq.\ 79.}]lindeberg:1994, klein:2007}
\begin{equation}
  \U = - \Minv\, \dbdtxn , \label{Ubasic}  
\end{equation}
which is valid for vector fields of arbitrary dimension.  This derivation provides an exact result as long as \M\ is non-singular.  

An alternate derivation for Eq.\ \ref{Ubasic} starts from the first order Taylor series expansion of \B\ with respect to time and space about a magnetic null point,
\begin{equation}
    \B \left( \delta \x, \delta t \right) = \M \cdot \delta \x + \dbdtxn \delta t
    + \mathcal{O}(\|\delta\x\|^2, \delta t^2)
    . \label{timedep}
\end{equation}
This first order expansion is valid in the limit of small $\delta t$ and $\left|\delta\x\right|$.  We define $\delta \xn$ as the position of the null point at $\delta t$.  Setting $\B \left( \delta \xn, \delta t \right)=0$ provides a unique solution for $\U \equiv \delta \xn / \delta t$, and we again arrive at Eq.\ \ref{Ubasic}.  Unlike the previous paragraph, this derivation uses the linearization approximation. Eq.\ \ref{Ubasic} may also be derived from the implicit function theorem. 

Equation \ref{Ubasic} shows that a null point will move along the path for which \B\ and \dbdtxn\ are oppositely directed.  The null point will move faster if the vector field is changing quickly in time or varying slowly in space along this path.  This exact result for \U\ can be applied to find the velocity of linear null points in any time-varying vector field with continuous first derivatives in time and space about the null point.  A unique velocity \U\ exists as long as \M\ is non-singular. If \M\ is non-singular, then there exists exactly one radial path away from the null for which the vector field is pointed in a particular direction.  

\section{Motion of Magnetic Null Points\label{magnetic}}

We next consider the case where \B\ is a magnetic field rather than just any vector field.  The derivation of Eq.\ \ref{Ubasic} does not invoke any of Maxwell's equations.  We now introduce Faraday's law,
\begin{equation}
 \dbdt = - \nabla \times \E, \label{faraday}
\end{equation}
where \E\ is the electric field.  By combining Eqs.\ \ref{Ubasic} and \ref{faraday}, we arrive at the relation 
\begin{equation}
  \U = \Minv \, (\nabla \times \E), \label{Ufaraday}
\end{equation}
which additionally requires continuous first derivatives of the electric field in space about the null point.  This expression does not depend on any particular Ohm's law, and indeed can be applied in situations where there is no Ohm's law.  

\begin{figure}
    \includegraphics[width=8.5cm]{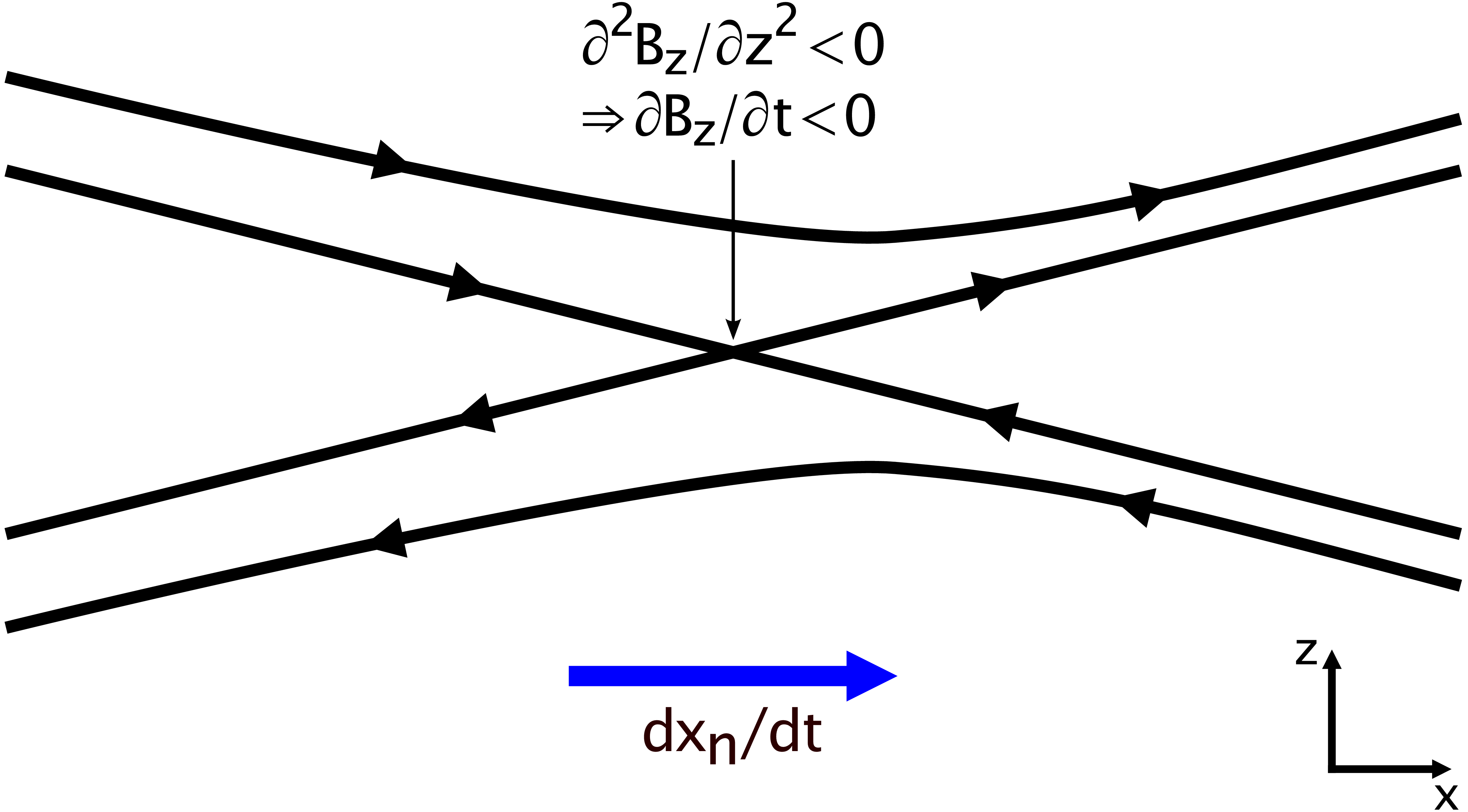}
    \caption{A two-dimensional example showing the motion of an X-type null point dominated by resistive diffusion of $B_z$ along the $z$ direction.  Above and below the null, $B_z<0$.  The negative $B_z$ diffuses along the $z$ direction into the immediate vicinity of the null point.  At a slightly later time, the magnetic field at the current position of the null point will have $B_z<0$.  The negative $B_z$ diffusion cancels out positive $B_z$ to the right of the null point, so the resulting null point motion is to the right.  Reproduced with permission from Ref.\ \onlinecite{murphy:retreat}. Copyright 2010 American Institute of Physics.}
        \label{mechanism}
\end{figure}

Next we consider the resistive MHD Ohm's law,
\begin{equation}
  \E + \V \times \B = \eta \J, \label{ohms_rmhd}
\end{equation}
where \V\ is the plasma flow velocity and \J\ is the current density.  The resistivity $\eta$ is assumed to be uniform for simplicity.  Eq.\ \ref{Ufaraday} then becomes
\begin{equation}
  \U = \V
  - \eta \Minv \nabla^2 \B, 
  \label{Urmhd}
\end{equation}
where all quantities on the right hand side are evaluated at the magnetic null.  This expression requires that \B\ has continuous first derivatives in time and continuous second derivatives in space about the null point.  Null point motion in resistive MHD results from a combination of advection by the bulk plasma flow and resistive diffusion of the magnetic field.  Even in the absence of flow, null points may still move in resistive situations.  The plasma flow velocity \emph{at} the null point does not equal the velocity of the null point itself.\cite{murphy:retreat}  A schematic showing null point motion due to resistive diffusion is presented in Fig.\ \ref{mechanism}.

Equation \ref{Ufaraday} can also be evaluated using an Ohm's law containing additional terms.  For example, we can choose our Ohm's law to be
\begin{equation}
    \E + \Vi \times \B = \eta \J 
    + \frac{\J\times\B}{en_e} 
    - \frac{\nabla p_e}{en_e} \label{hallohms},
\end{equation}
where $\V_i$ is the bulk ion velocity, $n_e$ is the electron density, $e$ is the elementary charge, and $p_e$ is a scalar electron pressure.  For $\J=en_e\left(\Vi - \Ve\right)$, Eq.\ \ref{Ufaraday} becomes
\begin{equation}
\U = \Ve - \eta\,\Minv\,\nabla^2\B + \Minv\left(\frac{\nabla n_e \times \nabla p_e}{n_e^2 e}\right)
      \label{Utwofl}
\end{equation}
where quantities are again evaluated at the null point. The first term on the right hand side corresponds to the magnetic field being carried with the electron flow velocity, \Ve, rather than the bulk plasma flow, the second term corresponds to the resistive diffusion of the magnetic field at the null, and the third term corresponds to the Biermann battery.

\section{The appearance and disappearance of magnetic null points\label{bifurcation}}

We next consider the emergence and disappearance of magnetic null points, with an emphasis on the instantaneous velocity of separation or convergence of the bifurcating null-null pair.  The local approach taken here complements global bifurcation studies.\cite{Priest:1996A, *DBrown:1999B, *DBrown:2001} Thus far we have only considered non-degenerate null points for which the local magnetic field can be described by Eq.\ \ref{BMr} using only the linear terms in the Taylor series expansion.  As long as \M\ is non-singular at the null, then there exists a unique velocity corresponding to the motion of that null point.  Non-degenerate null points are therefore structurally stable and cannot disappear unless \M\ becomes singular.\cite{greene:1993}

In contrast, degenerate null points are structurally unstable and generally exist instantaneously as a transition between different topological states.\cite{greene:1988, hornig:1996}  
Null points must appear or disappear in oppositely signed pairs during a bifurcation because the overall topological degree of the region cannot change unless a null point enters or leaves the domain across a boundary.\cite{deimling:1985} In most situations of physical interest, degenerate three-dimensional magnetic null points will have $\rank\M=2$ and $\nullity\M=1$ (e.g., Ref.\ \onlinecite{greene:1988}). The null space (or kernel) of \M\ will then be one-dimensional and corresponds to the eigenvector of \M\ with eigenvalue zero.  The three eigenvalues must sum to zero because of the divergence constraint,\cite{fukao:1975} which implies that the two non-zero eigenvalues must either be both real and of opposite sign, or both complex and of opposite sign.\cite{parnell:1996}

Although the linear representation in Eq.\ \ref{BMr} can describe the magnetic structure surrounding a degenerate null point (e.g., Ref.~\onlinecite{parnell:1996}) the region around a bifurcating null-null pair requires higher-order terms.  Third-order terms need to be considered only when the first- and second-order derivatives both vanish at the null, so usually a second-order expansion will suffice. The Taylor series expansion of the magnetic field about a three-dimensional null point to second-order in time and space is 
\begin{equation}
\begin{split}
    \B(\delta\x,\delta t)
     = 
    \M\cdot\delta\x
        + 
    \dbdtxn \delta t 
    + \frac{1}{2} 
    \left[
        \begin{array}{c}
        \delta\x^T \, \Hx \, \delta\x \\
        \delta\x^T \, \Hy \, \delta\x \\
        \delta\x^T \, \Hz \, \delta\x
        \end{array}
    \right] 
    \hspace{12mm}
    \\  
    \hspace{4mm}
    + \frac{\delta t}{2}
        \left[
            \begin{array}{c}
                (\delta\x\cdot\nabla ) \partial_t B_x \\
                (\delta\x\cdot\nabla ) \partial_t B_y \\
                (\delta\x\cdot\nabla ) \partial_t B_z \\
            \end{array}
        \right]
            + 
    \frac{\delta t^2}{2} \frac{\partial^2 \B_n}{\partial t^2}
    + \mathcal{O}(\|\delta\x\|^3,\delta t^3),
     \label{hessiantensor}
     \end{split}
\end{equation}
where the Jacobian matrix, Hessian matrices, and derivatives are evaluated at the null point.
The elements of the Hessian matrices are given by $\H_{k,ij} = \partial_i \partial_j B_k$ for $i,j,k \in \{x,y,z\}$.  If the magnetic field is locally continuous, then the partial derivative operators will be commutative and the Hessian matrices will be symmetric.  When $\dot{\B}$ is constant in time and space, then the fourth and fifth terms on the right hand side of Eq.\ \ref{hessiantensor} vanish. The positions of the null points for a given $\delta t$ may be found by setting $\B(\delta\x_n,\delta t) =0$ in Eq.~\ref{hessiantensor} (or the full expression of the magnetic field) and then solving for $\delta \x_n$.  

The instantaneous velocity of convergence or separation of a bifurcating null-null pair may be infinite, finite, or zero (see Section \ref{bifex} for examples).  Suppose that there exists a degenerate three-dimensional null point with $\rank{\M}=2$ and that \M\ has three unique eigenvectors.  The first-order directional derivatives of each component of \B\ are zero along the one-dimensional null space of \M\@.  Under most realistic circumstances, the second-order directional derivative of \B\ along the null space of \M\ will be nonzero.  Except in special circumstances, the component of velocity along the null space of \M\ of the bifurcating null-null pair will be instantaneously infinite.  Next, consider the two-dimensional subspace that is orthogonal to the null space of \M\@.  The Jacobian of \M\ in this subspace at the null point will be invertible.  Consequently, there exists a unique finite velocity within this subspace for the two-dimensional null point.  Thus, in general, the instantaneous component of velocity along the null space of a bifurcating null-null pair will be infinite while the components of velocity orthogonal to the null space will be finite or zero.

Next we consider separators that may exist and connect a bifurcating null-null pair.  Because these bifurcations cannot change the topological degree of the system, the null-null pair will include one negative null and one positive null.\cite{deimling:1985, greene:1988}  Define $\SigmaA$ and $\SigmaB$ as the separatrix surfaces and $\gammaA$ and $\gammaB$ as the spine field lines of the negative and positive null points, respectively.  The field lines in $\SigmaA$ and $\gammaB$ approach the null, while the field lines in $\SigmaB$ and $\gammaA$ recede from the null.  In the neighborhood of a linear null, the separatrix surface is given by the plane spanned by the two eigenvectors associated with eigenvalues that have the same sign for their real part, and the two spine field lines are along the remaining eigenvector.\cite{parnell:1996}

Separators that exist in real systems will almost always be given by the intersection of two separatrix surfaces.\cite{greene:1988, haynes:2010}  Spine-spine separators may exist if \gammaA\ and \gammaB\ include the same field line.  Though spine-spine separators may occur in some symmetric systems, they are not structurally stable and thus can generally be ignored.\cite{haynes:2010}  
As explained above, during the bifurcation of a degenerate null point a positive and negative null are formed; hence, spine-fan separators can never connect a bifurcating null-null pair because such separators connect either two positive or two negative nulls.  Additionally, not all bifurcating null-null pairs will be connected by a separator.  

In most realistic situations, there will not exist a straight line separator connecting a bifurcating null-null pair as one might intuitively expect (see also Refs.\ \onlinecite{greene:1988} and \onlinecite{lau:1990}).  Typically, there will exist some angle between the separatrix surfaces of each of the two null points in the time surrounding the bifurcation.  Equivalently, each pair of eigenvectors associated with the separatrix surface of each null will usually be changing in time, and, in general, this evolution will be different for each separatrix surface.

A straight line separator may only be created under special circumstances, such as when certain symmetries are present.  For example, a straight line separator will occur if the two nulls from the bifurcating null-null pair are both improper nulls (not spiral) and they both share the same fan eigenvector which is parallel to the direction of motion of the bifurcating nulls.

\subsection{Bifurcation Examples\label{bifex}}

\newcommand\nulldet{\ensuremath{a^2+bc}}
\newcommand\sqrtnulldet{\ensuremath{\sqrt{\nulldet}}}

\newcommand\eone{{\ensuremath{\mathbf{e}_1}}}
\newcommand\etwo{{\ensuremath{\mathbf{e}_2}}}

Let us consider a prototypical null point bifurcation of the form
\begin{equation}
    \B\left(\x,t\right) = 
        \begin{bmatrix}
            (a-z)x + by \\ cx - (a+z)y \\ z^2
        \end{bmatrix}
    + \delta\B(\x,t)
    , \label{bifurcationexample}
\end{equation}
where $a$, $b$, and $c$ are arbitrary real constants with $\nulldet \ne 0$. We assume that $\delta\B(\x,0)=0$ so that there exists a degenerate null point at the origin with $\rank\M=2$ at $t=0$.  The null space of \M\ at the degenerate null point is given by $\zhat$, which is the eigenvector of \M\ corresponding to eigenvalue zero.  The remaining eigenvectors of the degenerate null point are in the $x$-$y$ plane and given by 
\begin{equation}
        \eone \equiv 
            \begin{bmatrix}
                -b \\ a+\sqrtnulldet \\ 0
            \end{bmatrix} \mbox{and }
        \etwo \equiv 
            \begin{bmatrix}
                -b \\ a-\sqrtnulldet \\ 0
            \end{bmatrix} ,
\end{equation}
which correspond to eigenvalues $-(\nulldet)$ and $\nulldet$, respectively. 
We only consider time and space close to the bifurcation such that $\left|\delta B_z\right| < \left|\nulldet\right|$.  These examples will elucidate many of the properties of null point bifurcations discussed earlier in this section.

\subsubsection{First bifurcation example\label{bifexone}}

Suppose that $\delta\B(t)=-\zhat\sgn(t)\left|t\right|^\alpha$ in Eq.\ \ref{bifurcationexample}, with $\alpha>1$. 
For $t>0$, the third component of \B\ reduces to $z^2-t^\alpha$ and two null points exist, but when $t<0$, there are no null points.  At $t=0$, a single second-order null point appears at the origin as the system undergoes a saddle-node bifurcation.  For $t>0$, the two null points are at $\xn = \left[0,0,\pm t^{\alpha/2}\right]^\top$.  The null point with $z_n>0$ will be a positive null if $a^2+bc>0$ and a negative null otherwise.  The null points have velocities of $\dotxn = \left[0,0,\pm \frac{\alpha}{2}t^{\alpha/2-1} \right]^\top$.  When $1<\alpha<2$, the velocity of separation diverges to infinity at $t=0$.  For the critical case when $\alpha=2$, the null-point velocities are constant: $\dotxn = \left[0,0,\pm 1\right]^\top$.  When $\alpha>2$, the null-point velocities asymptotically approach zero at $t=0$.  Since the null point velocities are purely in the $\zhat$ direction and this is also the direction of an eigenvector of the fan planes of each of the nulls, the resulting separator is a straight line along $x=y=0$ for which $B_z<0$.  

Finely tuned examples such as this one are unlikely to occur in nature, but show that instantaneous velocities that are infinite, finite, or zero are mathematically allowable during the bifurcation of a degenerate null point.  

\subsubsection{Second bifurcation example\label{bifextwo}}

Suppose that $\delta\B(t)= \dot{\B}\hspace{0.1mm}t$ in Eq.\ \ref{bifurcationexample}, where $\dot{B}_x$, $\dot{B}_y$, and $\dot{B}_z$ are real constants with $\dot{B}_z<0$.  By setting $\B(\xn,t)=0$, we arrive at this expression for the null point positions for $t\geq 0$, 
\begin{equation}
    \mathbf{x}_{n}(t) =       
            \begin{bmatrix}  
                \dfrac{-a \dbx t - b \dby t \mp \dbx\sqrt{-\dbz}t^{3/2}}{\nulldet + \dbz t}
                \vspace{1.6mm}\\
                \dfrac{-c\dbx t + a\dby t \mp \dby \sqrt{-\dbz}t^{3/2}}{\nulldet + \dbz t}
                \vspace{1.6mm}\\
                \pm \sqrt{-\dbz t}
            \end{bmatrix}.
            \label{nullposbif}
\end{equation}
The instantaneous velocities of the bifurcating null points at $t=0$ are given by
\begin{equation}
    \lim_{t\rightarrow 0^+} \dot{\x}_{n}(t) = 
        \begin{bmatrix} 
            \dfrac{- a\dot{B}_x - b\dot{B}_y}{a^2+bc} 
            \vspace{1.6mm}\\
            \dfrac{- c\dot{B}_x + a\dot{B}_y}{a^2+bc} 
            \vspace{1.6mm}\\
            \pm \infty
        \end{bmatrix}.
\end{equation}
The velocity in the $x$-$y$ plane will be finite except under the special circumstance when $\dbx=\dby=0$ in which case the velocity in the $x$-$y$ plane will be zero.  The instantaneous component of velocity along the null space of \M\ is infinite.  

\begin{table*}
\caption{\label{eigentable}The eigenvectors, eigenvalues, and direction normal to the fan for the null points in the second bifurcation example.}
\begin{ruledtabular}
\begin{tabular}{lcccc}
 & \multicolumn{2}{c}{Case 1: $\nulldet>z_n^2>0$} & \multicolumn{2}{c}{Case 2: $\nulldet<0$} \\
 & Pos.\ null ($z_n>0$) & Neg.\ null ($z_n<0$) & Pos.\ null ($z_n<0$) & Neg.\ null ($z_n>0$) \\ \hline
Fan eigenvector, eigenvalue & \zhat, $2|z_n|$ & \zhat, $-2|z_n|$ & \eone, $|z_n|+\sqrtnulldet$ & \eone, $-|z_n|-\sqrtnulldet$ \\
Fan eigenvector, eigenvalue & \etwo, $-|z_n|+\sqrtnulldet$ & \eone, $|z_n|-\sqrtnulldet$  & \etwo, $|z_n|-\sqrtnulldet$ & \etwo, $-|z_n|+\sqrtnulldet$\\
Spine eigenvector, eigenvalue & \eone, $-|z_n|-\sqrtnulldet$ & \etwo, $|z_n|+\sqrtnulldet$ & \zhat, $-2|z_n|$ & \zhat, $2|z_n|$\\
Direction normal to fan & $[-a+\sqrtnulldet,-b,0]^\top$ & $[-a-\sqrtnulldet,-b,0]^\top$ & \zhat & \zhat 
\end{tabular}
\end{ruledtabular}
\end{table*}

\begin{figure}
   \includegraphics[width=8.5cm]{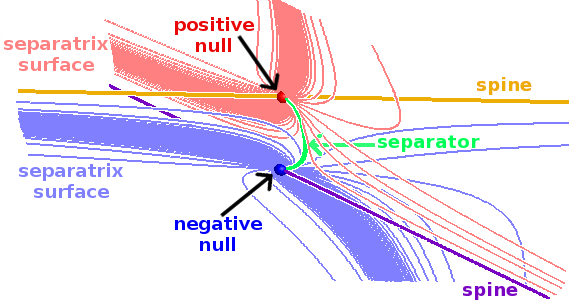}
    \caption{ Two improper null points (red and blue spheres) resulting from the bifurcation of a degenerate null point with $a^2+bc>z_n^2>0$.  The fan surfaces of the two nulls (denoted by salmon field lines for the positive null at $z_n>0$ and light blue field lines for the negative null at $z_n<0$) intersect to yield a curved separator field line (green).  The spine field lines are orange/purple for the positive/negative null.  In this example, 
    $(a,b,c)=(2,-1,3)$, $\dot{\mathbf{B}}=[1,-1,-1]^\top$, and $t=0.2$.
    }\label{realnulls}
\end{figure}

\begin{figure}
    \includegraphics[width=8.5cm]{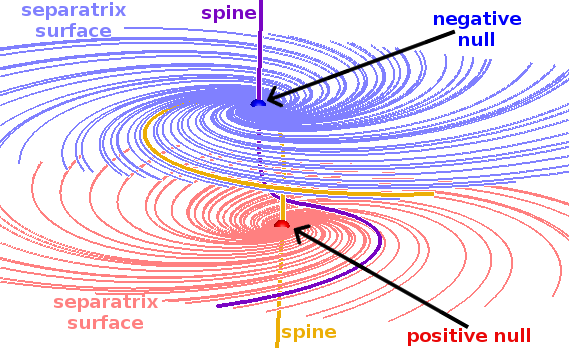}
    \caption{ Two spiral null points (red and blue spheres) resulting from the bifurcation of a degenerate null point with $a^2+bc<0$. The spine field lines (thick red/dark blue lines for the positive/negative nulls) wrap around each other instead of intersecting. The fan surfaces (salmon/light blue lines for the positive/negative nulls) are parallel to each other and the $x$-$y$ plane so do not intersect. Hence, no separator connects this bifurcating null-null pair.  In this example,
    $(a,b,c)=(1,1,-2)$, $\dot{\mathbf{B}}=[1,1,-2]^\top$, and $t=0.1$.
    }\label{spiralnulls_nosep}
\end{figure}

The eigenvectors, eigenvalues, and direction normal to the fan for the null points resulting from the bifurcation are shown in Table \ref{eigentable}.  The eigenvectors associated with each null are not functions of time for this example, but the eigenvalues are.  The structure of the resulting null-null pair depends on the value of \nulldet.  

When $\nulldet>z_n^2>0$ (Case 1), all eigenvalues are real so the bifurcation results in a positive improper null point with $z_n>0$ and a negative improper null point with $z_n<0$. The separatrix surfaces cannot be parallel in such a case, so a separator exists between the two nulls as the curved intersection of the two separatrix surfaces (see Figure~\ref{realnulls} for a typical example).

When $\nulldet<0$ (Case 2), each null has two complex conjugate eigenvalues so the bifurcation results in a positive spiral null point with $z_n<0$ and a negative spiral null point with $z_n>0$.  The field lines in the separatrix surfaces of both nulls are parallel to the $x$-$y$ plane, so the separatrix surfaces do not intersect to yield a separator.  A spine-spine separator can exist under special conditions (e.g., when $\dbx=\dby=0$), but under generic conditions no separator will exist to connect these two newly formed null points.  
Figure~\ref{spiralnulls_nosep} shows an example where the spine field lines of each null twist around each other before approaching the fan of the other null and spiraling away.  

A separator will exist as a straight line between the two bifurcating null points if and only if $\dbx=\dby=0$ for both Case 1 or Case 2\@.  

\subsubsection{Third bifurcation example\label{bifexthree}}

In contrast to the first two examples, we now consider a magnetic field perturbation that is a function of both time and space.  We define $\delta\B(\x,t) = \dot{\B}(\x)\hspace{0.1mm}t$ in Eq.\ \ref{bifurcationexample}, where \dbx\ and \dby\  are real constants and $\dbz(y)=3y-2$ is a linear function of $y$.  For the particular case shown in Figure 4, $(a,b,c)=(1,1,-2)$.  A separator is created connecting the bifurcating null-null pair because both the eigenvalues and eigenvectors of the nulls evolve in time.  The separators formed in the case between two bifurcating spiral nulls are typically long and highly spiralled.  Solving ${\mathbf{B}}=0$ using the above values in Eq.\ \ref{bifurcationexample} gives the following as the null point locations as a function of $t$,
\begin{equation}
    \mathbf{x}_{n}(t) =       
            \begin{bmatrix}  
                -\frac{1}{2}\left(1\pm\sqrt{2t-3ty_n}\right)y_n
                \vspace{0.6mm}
                \\
                y_n
                \vspace{0.6mm}
                \\
                \pm \sqrt{2t-3ty_n}
            \end{bmatrix},
            \label{nullposbif2}
\end{equation}
where $y_n = \left[\left(1+2t\right)+\sqrt{28t^2+4t+1}\right]/6t$.  In the limit as $t\rightarrow 0$, we must consider the quadratic equation satisfied by $y_n$,
\begin{equation}
    3ty_n^2 -(1+2t)y_n -2t = 0,
\end{equation}
which implies $y_n=0$ (and, hence, $x_n=z_n=0$) as $t\rightarrow 0$. 
Differentiating this quadratic with respect to $t$ and then taking the limit $t\rightarrow 0$ reveals that $\dot{y_n}=-2$. It can then be shown that in the limit $t\rightarrow0$, the instantaneous velocities of the bifurcating null points are $\dotxn=[1,-2,\infty]^\top$.

\begin{figure}
    \includegraphics[width=8.5cm]{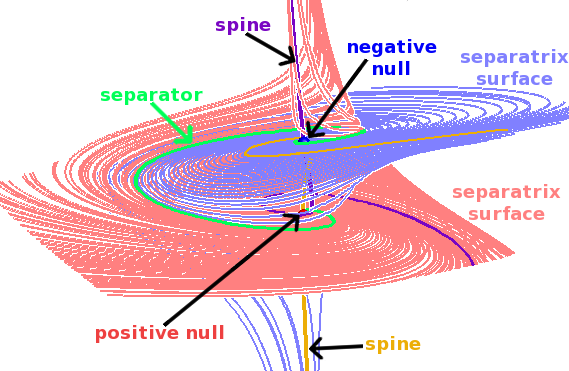}
    \caption{ As Figure~\ref{spiralnulls_nosep}, but here the time derivative of the $z$-component of the magnetic field, $\dot{B_z}$, is linear in $y$. The addition of a second-order term in space and time means that the fan surfaces (salmon/light blue lines for the positive/negative nulls) tilt towards each other the instant after bifurcation creating a separator which connects this bifurcating null-null pair.  In this example, $(a,b,c)=(1,1,-2)$, $t=0.1$, and $\delta{\B}(\x,t) = [t,0,3yt-2t]^\top$.
    }\label{spiralnulls_sep}
\end{figure}

\section{Discussion\label{discussion}}

In this paper, we derive an exact expression for the motion of linear null points in a vector field and apply this expression to magnetic null points.  Resistive diffusion and other effects in the generalized Ohm's law allow for non-ideal flows across magnetic null points.  In resistive MHD, null point motion results from a combination of advection by the bulk plasma flow and resistive diffusion of the magnetic field.  These results are particularly relevant to studies of null point magnetic reconnection, especially when asymmetries are present.  Analytical models of asymmetric reconnection must necessarily satisfy these expressions.  Non-ideal flows at null points allow the transfer of plasma across topological boundaries.

Just as we must be careful when describing the motion of magnetic field lines,\cite{newcomb:1958, *stern:1966, *vasyliunas:1972} we must also be careful when describing the motion of magnetic null points.  Null points are not objects.  A null point is not permanently affixed to a parcel of plasma except in ideal or certain perfectly symmetric cases.  Null points cannot be pushed directly by plasma pressure gradients or other forces on the plasma, but there will generally be indirect coupling between the momentum equation and Faraday's law that contributes to null-point motion.  The motion of a null point is determined intrinsically by local quantities evaluated at the null point.  However, global dynamics help set the local conditions that determine null-point motion.

In addition to providing insight into the physics of non-ideal flows at magnetic null points and constraining models of asymmetric reconnection, the expressions for null-point motion have several practical applications. Locating nulls of vector fields in three dimensions is non-trivial,\cite{greene:1992, *haynes:2007:trilinear} but if the null-point positions are found for one time, then these expressions provide a method for estimating the positions of null points at future times.  When there exists a cluster of several null points, these expressions provide a method for identifying which null points correspond to each other at different times.  A practical limitation is that these expressions will often require evaluating derivatives of noisy or numerical data (cf.\ Ref.\ \onlinecite{cullum:1971, *chartrand:2011}).  However, these expressions provide a test of numerical convergence and can be used to estimate the effective numerical resistivity in simulations of null-point reconnection (compare to Ref.\ \onlinecite{edmondson:2010a, *shen:2011}).

Linear magnetic null points appear and disappear in pairs associated with the bifurcation of a degenerate magnetic null point.  The null space of \M\ in these degenerate nulls will typically be one-dimensional.  Second or higher order terms in the Taylor series expansion are necessary to describe the structure of a degenerate null point and the region between a bifurcating null-null pair.  Except in special circumstances, the instantaneous velocity of convergence or separation of a null-null pair will typically be infinite along the null space of \M\ but with finite components of velocity in the orthogonal directions.  This means that null-null pairs that have just appeared or are just about to disappear will not lie next to one another, but will always have a finite separation no matter how small a time step between frames is taken and regardless of whether the field is known numerically on a grid or analytically everywhere within a domain.

Just before or after a bifurcation, a straight line separator connecting the null-null pair will generally not exist.  Furthermore, a separator, curved or straight, generic or non-generic, will not necessarily connect a null-null pair just before or after bifurcation if the nulls involved are of spiral type and their separatrix surfaces are parallel. The structures of second-order nulls and separators that exist very near a bifurcation remain important problems for future work. 

In resistive MHD, null points must resistively diffuse in and out of existence.  In the reference frame of the moving plasma, a necessary condition for a degenerate null point to form is that the resistive term in the induction equation, $\eta\nabla^2\B$, be antiparallel to the magnetic field at the location of the impending degenerate null.  This places physics-based geometric constraints on when and where bifurcations are allowed to happen.

We may consider whether or not a similar local analysis can be performed to describe the motion of separators.  Consider a separator that connects two magnetic null points.  Suppose that a segment of this separator exhibits non-ideal evolution.  Along the remainder of its length, the magnetic field in the vicinity of the separator evolves ideally.  At a slightly later time, the field line that was the separator will, in general, not continue to be the separator between these two null points despite the locally ideal evolution.  The motion of separators therefore cannot be described using solely local parameters.  However, it may be possible to derive an expression for the motion of a separator by taking into account plasma flow and connectivity changes along its entire length.  Such an approach would provide insight into the structural stability of separators and separator bifurcations,\cite{DBrown:1999A} as well as the nature of plasma flows across topological boundaries.  This latter aspect is fundamental to the basic physics of three-dimensional reconnection; indeed, an early definition\cite{vasyliunas:1975} states that reconnection is ``the process whereby plasma flows across a surface that separates regions containing topologically different magnetic field lines'' (see also Ref.\ \onlinecite{Schindler:1988}).  We are investigating the problem of separator motion in two and three dimensions in ongoing work.

There exist numerous additional opportunities for future work.  Our results take a local approach; consequently, numerical simulations are needed to investigate the interplay between local and global scales during null-point motion and bifurcations.  Numerical simulations can be used to investigate how null points diffuse in and out of existence in non-ideal plasmas and how separators behave during bifurcations.  If the flow field and magnetic field are well diagnosed in space or laboratory plasmas, the expressions for null-point motion may be used to provide constraints on magnetic field dissipation.  Equation \ref{Urmhd} offers another opportunity to measure the plasma resistivity in the collisional limit.  Dedicated laboratory experiments offer an opportunity to investigate plasma flow across null points (and other topological boundaries) as well as null point and separator bifurcations.  Finally, many results exist in the literature outside of plasma physics on bifurcations of vector fields and topology-based visualization of vector fields.  While communication across disciplines is hindered by differences in terminology,\footnote{Null points are also known as neutral points, fixed points, stationary points, equilibrium points, critical points, singular points, and singularities.  Separators are also known as saddle connectors and separation/attachment lines.  Separatrix surfaces are also known as fans and separation surfaces.} the application of this external knowledge to plasma physics will likely lead to improved physics-based understanding of these processes.

\begin{acknowledgments}

The authors thank A.\ Aggarwal, A.\ Bhattacharjee, A.\ Boozer, P.\ Cassak, J.\ Dorelli, T.\ Forbes, L.\ Guo, Y.-M.\ Huang, J.\ Lin, V.\ Lukin, M.\ Oka, E.\ Priest, J.\ Raymond, K.\ Reeves, C.\ Shen, V.\ Titov, D.\ Wendel, and E.\ Zweibel for helpful discussions.  The authors thank A.\ Wilmot-Smith and D.\ Pontin for a discussion that helped elucidate why the motion of separators cannot be described locally.  N.A.M.\ acknowledges support from NASA grants {{NNX11AB61G}}, {{NNX12AB25G}}, and {{NNX15AF43G}}; NASA contract {{NNM07AB07C}}; and NSF SHINE grants {{AGS-1156076}} and {{AGS-1358342}} to SAO\@. C.E.P.\ acknowledges support from the St Andrews 2013 STFC Consolidated grant.  This research has made use of NASA's Astrophysics Data System Bibliographic Services. The authors thank the journal for considering a manuscript containing only null results.

\end{acknowledgments}


\begin{thebibliography}{86}%
\makeatletter
\providecommand \@ifxundefined [1]{%
 \@ifx{#1\undefined}
}%
\providecommand \@ifnum [1]{%
 \ifnum #1\expandafter \@firstoftwo
 \else \expandafter \@secondoftwo
 \fi
}%
\providecommand \@ifx [1]{%
 \ifx #1\expandafter \@firstoftwo
 \else \expandafter \@secondoftwo
 \fi
}%
\providecommand \natexlab [1]{#1}%
\providecommand \enquote  [1]{``#1''}%
\providecommand \bibnamefont  [1]{#1}%
\providecommand \bibfnamefont [1]{#1}%
\providecommand \citenamefont [1]{#1}%
\providecommand \href@noop [0]{\@secondoftwo}%
\providecommand \href [0]{\begingroup \@sanitize@url \@href}%
\providecommand \@href[1]{\@@startlink{#1}\@@href}%
\providecommand \@@href[1]{\endgroup#1\@@endlink}%
\providecommand \@sanitize@url [0]{\catcode `\\12\catcode `\$12\catcode
  `\&12\catcode `\#12\catcode `\^12\catcode `\_12\catcode `\%12\relax}%
\providecommand \@@startlink[1]{}%
\providecommand \@@endlink[0]{}%
\providecommand \url  [0]{\begingroup\@sanitize@url \@url }%
\providecommand \@url [1]{\endgroup\@href {#1}{\urlprefix }}%
\providecommand \urlprefix  [0]{URL }%
\providecommand \Eprint [0]{\href }%
\providecommand \doibase [0]{http://dx.doi.org/}%
\providecommand \selectlanguage [0]{\@gobble}%
\providecommand \bibinfo  [0]{\@secondoftwo}%
\providecommand \bibfield  [0]{\@secondoftwo}%
\providecommand \translation [1]{[#1]}%
\providecommand \BibitemOpen [0]{}%
\providecommand \bibitemStop [0]{}%
\providecommand \bibitemNoStop [0]{.\EOS\space}%
\providecommand \EOS [0]{\spacefactor3000\relax}%
\providecommand \BibitemShut  [1]{\csname bibitem#1\endcsname}%
\let\auto@bib@innerbib\@empty
\bibitem [{\citenamefont {{Priest}}\ and\ \citenamefont
  {{Forbes}}(2000)}]{PF00}%
  \BibitemOpen
  \bibfield  {author} {\bibinfo {author} {\bibfnamefont {E.}~\bibnamefont
  {{Priest}}}\ and\ \bibinfo {author} {\bibfnamefont {T.}~\bibnamefont
  {{Forbes}}},\ }\href@noop {} {\emph {\bibinfo {title} {{Magnetic
  Reconnection: MHD Theory and Applications}}}}\ (\bibinfo  {publisher}
  {Cambridge University Press, Cambridge, UK},\ \bibinfo {year}
  {2000})\BibitemShut {NoStop}%
\bibitem [{\citenamefont {{Birn}}\ and\ \citenamefont
  {{Priest}}(2007)}]{birn:2007:book}%
  \BibitemOpen
  \bibinfo {editor} {\bibfnamefont {J.}~\bibnamefont {{Birn}}}\ and\ \bibinfo
  {editor} {\bibfnamefont {E.~R.}\ \bibnamefont {{Priest}}},\ eds.,\ \href@noop
  {} {\emph {\bibinfo {title} {Reconnection of magnetic fields:
  magnetohydrodynamics and collisionless theory and observations}}}\ (\bibinfo
  {publisher} {Cambridge University Press, Cambridge, UK},\ \bibinfo {year}
  {2007})\BibitemShut {NoStop}%
\bibitem [{\citenamefont {Zweibel}\ and\ \citenamefont
  {Yamada}(2009)}]{zweibel:2009}%
  \BibitemOpen
  \bibfield  {author} {\bibinfo {author} {\bibfnamefont {E.~G.}\ \bibnamefont
  {Zweibel}}\ and\ \bibinfo {author} {\bibfnamefont {M.}~\bibnamefont
  {Yamada}},\ }\bibfield  {title} {\enquote {\bibinfo {title} {{Magnetic
  Reconnection in Astrophysical and Laboratory Plasmas}},}\ }\href {\doibase
  10.1146/annurev-astro-082708-101726} {\bibfield  {journal} {\bibinfo
  {journal} {Annu. Rev. Astron. Astrophys.}\ }\textbf {\bibinfo {volume}
  {47}},\ \bibinfo {pages} {291--332} (\bibinfo {year} {2009})}\BibitemShut
  {NoStop}%
\bibitem [{\citenamefont {Yamada}, \citenamefont {Kulsrud},\ and\ \citenamefont
  {Ji}(2010)}]{yamada:2010}%
  \BibitemOpen
  \bibfield  {author} {\bibinfo {author} {\bibfnamefont {M.}~\bibnamefont
  {Yamada}}, \bibinfo {author} {\bibfnamefont {R.}~\bibnamefont {Kulsrud}}, \
  and\ \bibinfo {author} {\bibfnamefont {H.}~\bibnamefont {Ji}},\ }\bibfield
  {title} {\enquote {\bibinfo {title} {{Magnetic reconnection}},}\ }\href
  {\doibase 10.1103/RevModPhys.82.603} {\bibfield  {journal} {\bibinfo
  {journal} {Rev.\ Mod.\ Phys.}\ }\textbf {\bibinfo {volume} {82}},\ \bibinfo
  {pages} {603--664} (\bibinfo {year} {2010})}\BibitemShut {NoStop}%
\bibitem [{\citenamefont {{Cowley}}(1973)}]{cowley:1973}%
  \BibitemOpen
  \bibfield  {author} {\bibinfo {author} {\bibfnamefont {S.~W.~H.}\
  \bibnamefont {{Cowley}}},\ }\bibfield  {title} {\enquote {\bibinfo {title}
  {{A qualitative study of the reconnection between the Earth's magnetic field
  and an interplanetary field of arbitrary orientation}},}\ }\href {\doibase
  10.1029/RS008i011p00903} {\bibfield  {journal} {\bibinfo  {journal} {Radio
  Science}\ }\textbf {\bibinfo {volume} {8}},\ \bibinfo {pages} {903--913}
  (\bibinfo {year} {1973})}\BibitemShut {NoStop}%
\bibitem [{\citenamefont {Fukao}, \citenamefont {Ugai},\ and\ \citenamefont
  {Tsuda}(1975)}]{fukao:1975}%
  \BibitemOpen
  \bibfield  {author} {\bibinfo {author} {\bibfnamefont {S.}~\bibnamefont
  {Fukao}}, \bibinfo {author} {\bibfnamefont {M.}~\bibnamefont {Ugai}}, \ and\
  \bibinfo {author} {\bibfnamefont {T.}~\bibnamefont {Tsuda}},\ }\bibfield
  {title} {\enquote {\bibinfo {title} {{Topological study of magnetic field
  near a neutral point}},}\ }\href@noop {} {\bibfield  {journal} {\bibinfo
  {journal} {RISRJ}\ }\textbf {\bibinfo {volume} {29}},\ \bibinfo {pages}
  {133--139} (\bibinfo {year} {1975})}\BibitemShut {NoStop}%
\bibitem [{\citenamefont {Greene}(1988)}]{greene:1988}%
  \BibitemOpen
  \bibfield  {author} {\bibinfo {author} {\bibfnamefont {J.~M.}\ \bibnamefont
  {Greene}},\ }\bibfield  {title} {\enquote {\bibinfo {title} {{Geometrical
  properties of three-dimensional reconnecting magnetic fields with nulls}},}\
  }\href {\doibase 10.1029/JA093iA08p08583} {\bibfield  {journal} {\bibinfo
  {journal} {J.\ Geophys.\ Res.}\ }\textbf {\bibinfo {volume} {93}},\ \bibinfo
  {pages} {8583--8590} (\bibinfo {year} {1988})}\BibitemShut {NoStop}%
\bibitem [{\citenamefont {Lau}\ and\ \citenamefont {Finn}(1990)}]{lau:1990}%
  \BibitemOpen
  \bibfield  {author} {\bibinfo {author} {\bibfnamefont {Y.-T.}\ \bibnamefont
  {Lau}}\ and\ \bibinfo {author} {\bibfnamefont {J.~M.}\ \bibnamefont {Finn}},\
  }\bibfield  {title} {\enquote {\bibinfo {title} {{Three-dimensional kinematic
  reconnection in the presence of field nulls and closed field lines}},}\
  }\href {\doibase 10.1086/168419} {\bibfield  {journal} {\bibinfo  {journal}
  {Astrophys. J.}\ }\textbf {\bibinfo {volume} {350}},\ \bibinfo {pages}
  {672--691} (\bibinfo {year} {1990})}\BibitemShut {NoStop}%
\bibitem [{\citenamefont {Parnell}\ \emph {et~al.}(1996)\citenamefont
  {Parnell}, \citenamefont {Smith}, \citenamefont {Neukirch},\ and\
  \citenamefont {Priest}}]{parnell:1996}%
  \BibitemOpen
  \bibfield  {author} {\bibinfo {author} {\bibfnamefont {C.~E.}\ \bibnamefont
  {Parnell}}, \bibinfo {author} {\bibfnamefont {J.~M.}\ \bibnamefont {Smith}},
  \bibinfo {author} {\bibfnamefont {T.}~\bibnamefont {Neukirch}}, \ and\
  \bibinfo {author} {\bibfnamefont {E.~R.}\ \bibnamefont {Priest}},\ }\bibfield
   {title} {\enquote {\bibinfo {title} {{The structure of three-dimensional
  magnetic neutral points}},}\ }\href {\doibase 10.1063/1.871810} {\bibfield
  {journal} {\bibinfo  {journal} {Phys.\ Plasmas}\ }\textbf {\bibinfo {volume}
  {3}},\ \bibinfo {pages} {759--770} (\bibinfo {year} {1996})}\BibitemShut
  {NoStop}%
\bibitem [{\citenamefont {Xiao}\ \emph {et~al.}(2006)\citenamefont {Xiao},
  \citenamefont {Wang}, \citenamefont {Pu}, \citenamefont {Zhao}, \citenamefont
  {Wang}, \citenamefont {Ma}, \citenamefont {Fu}, \citenamefont {Kivelson},
  \citenamefont {Liu}, \citenamefont {Zong}, \citenamefont {Glassmeier},
  \citenamefont {Balogh}, \citenamefont {Korth}, \citenamefont {Reme},\ and\
  \citenamefont {Escoubet}}]{xiao:2006}%
  \BibitemOpen
  \bibfield  {author} {\bibinfo {author} {\bibfnamefont {C.~J.}\ \bibnamefont
  {Xiao}}, \bibinfo {author} {\bibfnamefont {X.~G.}\ \bibnamefont {Wang}},
  \bibinfo {author} {\bibfnamefont {Z.~Y.}\ \bibnamefont {Pu}}, \bibinfo
  {author} {\bibfnamefont {H.}~\bibnamefont {Zhao}}, \bibinfo {author}
  {\bibfnamefont {J.~X.}\ \bibnamefont {Wang}}, \bibinfo {author}
  {\bibfnamefont {Z.~W.}\ \bibnamefont {Ma}}, \bibinfo {author} {\bibfnamefont
  {S.~Y.}\ \bibnamefont {Fu}}, \bibinfo {author} {\bibfnamefont {M.~G.}\
  \bibnamefont {Kivelson}}, \bibinfo {author} {\bibfnamefont {Z.~X.}\
  \bibnamefont {Liu}}, \bibinfo {author} {\bibfnamefont {Q.~G.}\ \bibnamefont
  {Zong}}, \bibinfo {author} {\bibfnamefont {K.~H.}\ \bibnamefont
  {Glassmeier}}, \bibinfo {author} {\bibfnamefont {A.}~\bibnamefont {Balogh}},
  \bibinfo {author} {\bibfnamefont {A.}~\bibnamefont {Korth}}, \bibinfo
  {author} {\bibfnamefont {H.}~\bibnamefont {Reme}}, \ and\ \bibinfo {author}
  {\bibfnamefont {C.~P.}\ \bibnamefont {Escoubet}},\ }\bibfield  {title}
  {\enquote {\bibinfo {title} {{In situ evidence for the structure of the
  magnetic null in a 3D reconnection event in the Earth's magnetotail}},}\
  }\href {\doibase 10.1038/nphys342} {\bibfield  {journal} {\bibinfo  {journal}
  {Nature Physics}\ }\textbf {\bibinfo {volume} {2}},\ \bibinfo {pages}
  {478--483} (\bibinfo {year} {2006})}\BibitemShut {NoStop}%
\bibitem [{\citenamefont {{Xiao}}\ \emph {et~al.}(2007)\citenamefont {{Xiao}},
  \citenamefont {{Wang}}, \citenamefont {{Pu}}, \citenamefont {{Ma}},
  \citenamefont {{Zhao}}, \citenamefont {{Zhou}}, \citenamefont {{Wang}},
  \citenamefont {{Kivelson}}, \citenamefont {{Fu}}, \citenamefont {{Liu}},
  \citenamefont {{Zong}}, \citenamefont {{Dunlop}}, \citenamefont
  {{Glassmeier}}, \citenamefont {{Lucek}}, \citenamefont {{Reme}},
  \citenamefont {{Dandouras}},\ and\ \citenamefont {{Escoubet}}}]{xiao:2007}%
  \BibitemOpen
  \bibfield  {author} {\bibinfo {author} {\bibfnamefont {C.~J.}\ \bibnamefont
  {{Xiao}}}, \bibinfo {author} {\bibfnamefont {X.~G.}\ \bibnamefont {{Wang}}},
  \bibinfo {author} {\bibfnamefont {Z.~Y.}\ \bibnamefont {{Pu}}}, \bibinfo
  {author} {\bibfnamefont {Z.~W.}\ \bibnamefont {{Ma}}}, \bibinfo {author}
  {\bibfnamefont {H.}~\bibnamefont {{Zhao}}}, \bibinfo {author} {\bibfnamefont
  {G.~P.}\ \bibnamefont {{Zhou}}}, \bibinfo {author} {\bibfnamefont {J.~X.}\
  \bibnamefont {{Wang}}}, \bibinfo {author} {\bibfnamefont {M.~G.}\
  \bibnamefont {{Kivelson}}}, \bibinfo {author} {\bibfnamefont {S.~Y.}\
  \bibnamefont {{Fu}}}, \bibinfo {author} {\bibfnamefont {Z.~X.}\ \bibnamefont
  {{Liu}}}, \bibinfo {author} {\bibfnamefont {Q.~G.}\ \bibnamefont {{Zong}}},
  \bibinfo {author} {\bibfnamefont {M.~W.}\ \bibnamefont {{Dunlop}}}, \bibinfo
  {author} {\bibfnamefont {K.-H.}\ \bibnamefont {{Glassmeier}}}, \bibinfo
  {author} {\bibfnamefont {E.}~\bibnamefont {{Lucek}}}, \bibinfo {author}
  {\bibfnamefont {H.}~\bibnamefont {{Reme}}}, \bibinfo {author} {\bibfnamefont
  {I.}~\bibnamefont {{Dandouras}}}, \ and\ \bibinfo {author} {\bibfnamefont
  {C.~P.}\ \bibnamefont {{Escoubet}}},\ }\bibfield  {title} {\enquote {\bibinfo
  {title} {{Satellite observations of separator-line geometry of
  three-dimensional magnetic reconnection}},}\ }\href {\doibase
  10.1038/nphys650} {\bibfield  {journal} {\bibinfo  {journal} {Nature
  Physics}\ }\textbf {\bibinfo {volume} {3}},\ \bibinfo {pages} {609--613}
  (\bibinfo {year} {2007})}\BibitemShut {NoStop}%
\bibitem [{\citenamefont {{He}}\ \emph {et~al.}(2008)\citenamefont {{He}},
  \citenamefont {{Tu}}, \citenamefont {{Tian}}, \citenamefont {{Xiao}},
  \citenamefont {{Wang}}, \citenamefont {{Pu}}, \citenamefont {{Ma}},
  \citenamefont {{Dunlop}}, \citenamefont {{Zhao}}, \citenamefont {{Zhou}},
  \citenamefont {{Wang}}, \citenamefont {{Fu}}, \citenamefont {{Liu}},
  \citenamefont {{Zong}}, \citenamefont {{Glassmeier}}, \citenamefont {{Reme}},
  \citenamefont {{Dandouras}},\ and\ \citenamefont {{Escoubet}}}]{he:2008}%
  \BibitemOpen
  \bibfield  {author} {\bibinfo {author} {\bibfnamefont {J.-S.}\ \bibnamefont
  {{He}}}, \bibinfo {author} {\bibfnamefont {C.-Y.}\ \bibnamefont {{Tu}}},
  \bibinfo {author} {\bibfnamefont {H.}~\bibnamefont {{Tian}}}, \bibinfo
  {author} {\bibfnamefont {C.-J.}\ \bibnamefont {{Xiao}}}, \bibinfo {author}
  {\bibfnamefont {X.-G.}\ \bibnamefont {{Wang}}}, \bibinfo {author}
  {\bibfnamefont {Z.-Y.}\ \bibnamefont {{Pu}}}, \bibinfo {author}
  {\bibfnamefont {Z.-W.}\ \bibnamefont {{Ma}}}, \bibinfo {author}
  {\bibfnamefont {M.~W.}\ \bibnamefont {{Dunlop}}}, \bibinfo {author}
  {\bibfnamefont {H.}~\bibnamefont {{Zhao}}}, \bibinfo {author} {\bibfnamefont
  {G.-P.}\ \bibnamefont {{Zhou}}}, \bibinfo {author} {\bibfnamefont {J.-X.}\
  \bibnamefont {{Wang}}}, \bibinfo {author} {\bibfnamefont {S.-Y.}\
  \bibnamefont {{Fu}}}, \bibinfo {author} {\bibfnamefont {Z.-X.}\ \bibnamefont
  {{Liu}}}, \bibinfo {author} {\bibfnamefont {Q.-G.}\ \bibnamefont {{Zong}}},
  \bibinfo {author} {\bibfnamefont {K.-H.}\ \bibnamefont {{Glassmeier}}},
  \bibinfo {author} {\bibfnamefont {H.}~\bibnamefont {{Reme}}}, \bibinfo
  {author} {\bibfnamefont {I.}~\bibnamefont {{Dandouras}}}, \ and\ \bibinfo
  {author} {\bibfnamefont {C.~P.}\ \bibnamefont {{Escoubet}}},\ }\bibfield
  {title} {\enquote {\bibinfo {title} {{A magnetic null geometry reconstructed
  from Cluster spacecraft observations}},}\ }\href {\doibase
  10.1029/2007JA012609} {\bibfield  {journal} {\bibinfo  {journal} {J.\
  Geophys.\ Res.}\ }\textbf {\bibinfo {volume} {113}},\ \bibinfo {eid} {A05205}
  (\bibinfo {year} {2008})}\BibitemShut {NoStop}%
\bibitem [{\citenamefont {{Wendel}}\ and\ \citenamefont
  {{Adrian}}(2013)}]{wendel:2013}%
  \BibitemOpen
  \bibfield  {author} {\bibinfo {author} {\bibfnamefont {D.~E.}\ \bibnamefont
  {{Wendel}}}\ and\ \bibinfo {author} {\bibfnamefont {M.~L.}\ \bibnamefont
  {{Adrian}}},\ }\bibfield  {title} {\enquote {\bibinfo {title} {{Current
  structure and nonideal behavior at magnetic null points in the turbulent
  magnetosheath}},}\ }\href {\doibase 10.1002/jgra.50234} {\bibfield  {journal}
  {\bibinfo  {journal} {J.\ Geophys.\ Res.}\ }\textbf {\bibinfo {volume}
  {118}},\ \bibinfo {pages} {1571--1588} (\bibinfo {year} {2013})}\BibitemShut
  {NoStop}%
\bibitem [{\citenamefont {{Guo}}\ \emph {et~al.}(2013)\citenamefont {{Guo}},
  \citenamefont {{Pu}}, \citenamefont {{Xiao}}, \citenamefont {{Wang}},
  \citenamefont {{Fu}}, \citenamefont {{Xie}}, \citenamefont {{Zong}},
  \citenamefont {{He}}, \citenamefont {{Yao}}, \citenamefont {{Zhong}},\ and\
  \citenamefont {{Li}}}]{rguo:2013}%
  \BibitemOpen
  \bibfield  {author} {\bibinfo {author} {\bibfnamefont {R.}~\bibnamefont
  {{Guo}}}, \bibinfo {author} {\bibfnamefont {Z.}~\bibnamefont {{Pu}}},
  \bibinfo {author} {\bibfnamefont {C.}~\bibnamefont {{Xiao}}}, \bibinfo
  {author} {\bibfnamefont {X.}~\bibnamefont {{Wang}}}, \bibinfo {author}
  {\bibfnamefont {S.}~\bibnamefont {{Fu}}}, \bibinfo {author} {\bibfnamefont
  {L.}~\bibnamefont {{Xie}}}, \bibinfo {author} {\bibfnamefont
  {Q.}~\bibnamefont {{Zong}}}, \bibinfo {author} {\bibfnamefont
  {J.}~\bibnamefont {{He}}}, \bibinfo {author} {\bibfnamefont {Z.}~\bibnamefont
  {{Yao}}}, \bibinfo {author} {\bibfnamefont {J.}~\bibnamefont {{Zhong}}}, \
  and\ \bibinfo {author} {\bibfnamefont {J.}~\bibnamefont {{Li}}},\ }\bibfield
  {title} {\enquote {\bibinfo {title} {{Separator reconnection with
  antiparallel/component features observed in magnetotail plasmas}},}\ }\href
  {\doibase 10.1002/jgra.50569} {\bibfield  {journal} {\bibinfo  {journal} {J.\
  Geophys.\ Res.}\ }\textbf {\bibinfo {volume} {118}},\ \bibinfo {pages}
  {6116--6126} (\bibinfo {year} {2013})}\BibitemShut {NoStop}%
\bibitem [{\citenamefont {{Fu}}\ \emph {et~al.}(2015)\citenamefont {{Fu}},
  \citenamefont {{Vaivads}}, \citenamefont {{Khotyaintsev}}, \citenamefont
  {{Olshevsky}}, \citenamefont {{Andr{\'e}}}, \citenamefont {{Cao}},
  \citenamefont {{Huang}}, \citenamefont {{Retin{\`o}}},\ and\ \citenamefont
  {{Lapenta}}}]{fu:2015}%
  \BibitemOpen
  \bibfield  {author} {\bibinfo {author} {\bibfnamefont {H.~S.}\ \bibnamefont
  {{Fu}}}, \bibinfo {author} {\bibfnamefont {A.}~\bibnamefont {{Vaivads}}},
  \bibinfo {author} {\bibfnamefont {Y.~V.}\ \bibnamefont {{Khotyaintsev}}},
  \bibinfo {author} {\bibfnamefont {V.}~\bibnamefont {{Olshevsky}}}, \bibinfo
  {author} {\bibfnamefont {M.}~\bibnamefont {{Andr{\'e}}}}, \bibinfo {author}
  {\bibfnamefont {J.~B.}\ \bibnamefont {{Cao}}}, \bibinfo {author}
  {\bibfnamefont {S.~Y.}\ \bibnamefont {{Huang}}}, \bibinfo {author}
  {\bibfnamefont {A.}~\bibnamefont {{Retin{\`o}}}}, \ and\ \bibinfo {author}
  {\bibfnamefont {G.}~\bibnamefont {{Lapenta}}},\ }\bibfield  {title} {\enquote
  {\bibinfo {title} {{How to find magnetic nulls and reconstruct field topology
  with MMS data?}}}\ }\href {\doibase 10.1002/2015JA021082} {\bibfield
  {journal} {\bibinfo  {journal} {J.\ Geophys.\ Res.}\ }\textbf {\bibinfo
  {volume} {120}},\ \bibinfo {pages} {3758--3782} (\bibinfo {year}
  {2015})}\BibitemShut {NoStop}%
\bibitem [{\citenamefont {{Olshevsky}}\ \emph {et~al.}(2015)\citenamefont
  {{Olshevsky}}, \citenamefont {{Divin}}, \citenamefont {{Eriksson}},
  \citenamefont {{Markidis}},\ and\ \citenamefont
  {{Lapenta}}}]{olshevsky:2015}%
  \BibitemOpen
  \bibfield  {author} {\bibinfo {author} {\bibfnamefont {V.}~\bibnamefont
  {{Olshevsky}}}, \bibinfo {author} {\bibfnamefont {A.}~\bibnamefont
  {{Divin}}}, \bibinfo {author} {\bibfnamefont {E.}~\bibnamefont {{Eriksson}}},
  \bibinfo {author} {\bibfnamefont {S.}~\bibnamefont {{Markidis}}}, \ and\
  \bibinfo {author} {\bibfnamefont {G.}~\bibnamefont {{Lapenta}}},\ }\bibfield
  {title} {\enquote {\bibinfo {title} {{Energy Dissipation in Magnetic Null
  Points at Kinetic Scales}},}\ }\href {\doibase 10.1088/0004-637X/807/2/155}
  {\bibfield  {journal} {\bibinfo  {journal} {Astrophys. J.}\ }\textbf
  {\bibinfo {volume} {807}},\ \bibinfo {eid} {155} (\bibinfo {year}
  {2015})}\BibitemShut {NoStop}%
\bibitem [{\citenamefont {{Filippov}}(1999)}]{filippov:1999}%
  \BibitemOpen
  \bibfield  {author} {\bibinfo {author} {\bibfnamefont {B.}~\bibnamefont
  {{Filippov}}},\ }\bibfield  {title} {\enquote {\bibinfo {title} {{Observation
  of a 3D Magnetic Null Point in the Solar Corona}},}\ }\href {\doibase
  10.1023/A:1005124915577} {\bibfield  {journal} {\bibinfo  {journal} {Sol.
  Phys.}\ }\textbf {\bibinfo {volume} {185}},\ \bibinfo {pages} {297--309}
  (\bibinfo {year} {1999})}\BibitemShut {NoStop}%
\bibitem [{\citenamefont {{Aulanier}}\ \emph {et~al.}(2000)\citenamefont
  {{Aulanier}}, \citenamefont {{DeLuca}}, \citenamefont {{Antiochos}},
  \citenamefont {{McMullen}},\ and\ \citenamefont {{Golub}}}]{aulanier:2000}%
  \BibitemOpen
  \bibfield  {author} {\bibinfo {author} {\bibfnamefont {G.}~\bibnamefont
  {{Aulanier}}}, \bibinfo {author} {\bibfnamefont {E.~E.}\ \bibnamefont
  {{DeLuca}}}, \bibinfo {author} {\bibfnamefont {S.~K.}\ \bibnamefont
  {{Antiochos}}}, \bibinfo {author} {\bibfnamefont {R.~A.}\ \bibnamefont
  {{McMullen}}}, \ and\ \bibinfo {author} {\bibfnamefont {L.}~\bibnamefont
  {{Golub}}},\ }\bibfield  {title} {\enquote {\bibinfo {title} {{The Topology
  and Evolution of the Bastille Day Flare}},}\ }\href {\doibase 10.1086/309376}
  {\bibfield  {journal} {\bibinfo  {journal} {Astrophys. J.}\ }\textbf
  {\bibinfo {volume} {540}},\ \bibinfo {pages} {1126--1142} (\bibinfo {year}
  {2000})}\BibitemShut {NoStop}%
\bibitem [{\citenamefont {{Zhao}}\ \emph {et~al.}(2008)\citenamefont {{Zhao}},
  \citenamefont {{Wang}}, \citenamefont {{Zhang}}, \citenamefont {{Xiao}},\
  and\ \citenamefont {{Wang}}}]{zhao:2008}%
  \BibitemOpen
  \bibfield  {author} {\bibinfo {author} {\bibfnamefont {H.}~\bibnamefont
  {{Zhao}}}, \bibinfo {author} {\bibfnamefont {J.-X.}\ \bibnamefont {{Wang}}},
  \bibinfo {author} {\bibfnamefont {J.}~\bibnamefont {{Zhang}}}, \bibinfo
  {author} {\bibfnamefont {C.-J.}\ \bibnamefont {{Xiao}}}, \ and\ \bibinfo
  {author} {\bibfnamefont {H.-M.}\ \bibnamefont {{Wang}}},\ }\bibfield  {title}
  {\enquote {\bibinfo {title} {{Determination of the Topology Skeleton of
  Magnetic Fields in a Solar Active Region}},}\ }\href {\doibase
  10.1088/1009-9271/8/2/01} {\bibfield  {journal} {\bibinfo  {journal} {CJAA}\
  }\textbf {\bibinfo {volume} {8}},\ \bibinfo {pages} {133--145} (\bibinfo
  {year} {2008})}\BibitemShut {NoStop}%
\bibitem [{\citenamefont {{Longcope}}\ and\ \citenamefont
  {{Parnell}}(2009)}]{longcope:2009}%
  \BibitemOpen
  \bibfield  {author} {\bibinfo {author} {\bibfnamefont {D.~W.}\ \bibnamefont
  {{Longcope}}}\ and\ \bibinfo {author} {\bibfnamefont {C.~E.}\ \bibnamefont
  {{Parnell}}},\ }\bibfield  {title} {\enquote {\bibinfo {title} {{The Number
  of Magnetic Null Points in the Quiet Sun Corona}},}\ }\href {\doibase
  10.1007/s11207-008-9281-x} {\bibfield  {journal} {\bibinfo  {journal} {Sol.
  Phys.}\ }\textbf {\bibinfo {volume} {254}},\ \bibinfo {pages} {51--75}
  (\bibinfo {year} {2009})}\BibitemShut {NoStop}%
\bibitem [{\citenamefont {{Freed}}, \citenamefont {{Longcope}},\ and\
  \citenamefont {{McKenzie}}(2015)}]{freed:2015}%
  \BibitemOpen
  \bibfield  {author} {\bibinfo {author} {\bibfnamefont {M.~S.}\ \bibnamefont
  {{Freed}}}, \bibinfo {author} {\bibfnamefont {D.~W.}\ \bibnamefont
  {{Longcope}}}, \ and\ \bibinfo {author} {\bibfnamefont {D.~E.}\ \bibnamefont
  {{McKenzie}}},\ }\bibfield  {title} {\enquote {\bibinfo {title} {{Three-Year
  Global Survey of Coronal Null Points from Potential-Field-Source-Surface
  (PFSS) Modeling and Solar Dynamics Observatory (SDO) Observations}},}\ }\href
  {\doibase 10.1007/s11207-014-0616-5} {\bibfield  {journal} {\bibinfo
  {journal} {Sol. Phys.}\ }\textbf {\bibinfo {volume} {290}},\ \bibinfo {pages}
  {467--490} (\bibinfo {year} {2015})}\BibitemShut {NoStop}%
\bibitem [{\citenamefont {{Edwards}}\ and\ \citenamefont
  {{Parnell}}(2015)}]{edwards:2015}%
  \BibitemOpen
  \bibfield  {author} {\bibinfo {author} {\bibfnamefont {S.~J.}\ \bibnamefont
  {{Edwards}}}\ and\ \bibinfo {author} {\bibfnamefont {C.~E.}\ \bibnamefont
  {{Parnell}}},\ }\bibfield  {title} {\enquote {\bibinfo {title} {{Null Point
  Distribution in Global Coronal Potential Field Extrapolations}},}\ }\href
  {\doibase 10.1007/s11207-015-0727-7} {\bibfield  {journal} {\bibinfo
  {journal} {Sol. Phys.}\ } (\bibinfo {year} {2015}),\
  10.1007/s11207-015-0727-7}\BibitemShut {NoStop}%
\bibitem [{\citenamefont {{Masson}}\ \emph {et~al.}(2009)\citenamefont
  {{Masson}}, \citenamefont {{Pariat}}, \citenamefont {{Aulanier}},\ and\
  \citenamefont {{Schrijver}}}]{masson:2009}%
  \BibitemOpen
  \bibfield  {author} {\bibinfo {author} {\bibfnamefont {S.}~\bibnamefont
  {{Masson}}}, \bibinfo {author} {\bibfnamefont {E.}~\bibnamefont {{Pariat}}},
  \bibinfo {author} {\bibfnamefont {G.}~\bibnamefont {{Aulanier}}}, \ and\
  \bibinfo {author} {\bibfnamefont {C.~J.}\ \bibnamefont {{Schrijver}}},\
  }\bibfield  {title} {\enquote {\bibinfo {title} {{The Nature of Flare Ribbons
  in Coronal Null-Point Topology}},}\ }\href {\doibase
  10.1088/0004-637X/700/1/559} {\bibfield  {journal} {\bibinfo  {journal}
  {Astrophys. J.}\ }\textbf {\bibinfo {volume} {700}},\ \bibinfo {pages}
  {559--578} (\bibinfo {year} {2009})}\BibitemShut {NoStop}%
\bibitem [{\citenamefont {Demoulin}, \citenamefont {Henoux},\ and\
  \citenamefont {Mandrini}(1994)}]{demoulin:1994}%
  \BibitemOpen
  \bibfield  {author} {\bibinfo {author} {\bibfnamefont {P.}~\bibnamefont
  {Demoulin}}, \bibinfo {author} {\bibfnamefont {J.~C.}\ \bibnamefont
  {Henoux}}, \ and\ \bibinfo {author} {\bibfnamefont {C.~H.}\ \bibnamefont
  {Mandrini}},\ }\bibfield  {title} {\enquote {\bibinfo {title} {{Are magnetic
  null points important in solar flares?}}}\ }\href@noop {} {\bibfield
  {journal} {\bibinfo  {journal} {Astron. Astrophys.}\ }\textbf {\bibinfo
  {volume} {285}},\ \bibinfo {pages} {1023--1037} (\bibinfo {year}
  {1994})}\BibitemShut {NoStop}%
\bibitem [{\citenamefont {{Barnes}}(2007)}]{barnes:2007}%
  \BibitemOpen
  \bibfield  {author} {\bibinfo {author} {\bibfnamefont {G.}~\bibnamefont
  {{Barnes}}},\ }\bibfield  {title} {\enquote {\bibinfo {title} {{On the
  Relationship between Coronal Magnetic Null Points and Solar Eruptive
  Events}},}\ }\href {\doibase 10.1086/524107} {\bibfield  {journal} {\bibinfo
  {journal} {Astrophys. J. Lett.}\ }\textbf {\bibinfo {volume} {670}},\
  \bibinfo {pages} {L53--L56} (\bibinfo {year} {2007})}\BibitemShut {NoStop}%
\bibitem [{\citenamefont {{Close}}, \citenamefont {{Parnell}},\ and\
  \citenamefont {{Priest}}(2004)}]{Close:2004}%
  \BibitemOpen
  \bibfield  {author} {\bibinfo {author} {\bibfnamefont {R.~M.}\ \bibnamefont
  {{Close}}}, \bibinfo {author} {\bibfnamefont {C.~E.}\ \bibnamefont
  {{Parnell}}}, \ and\ \bibinfo {author} {\bibfnamefont {E.~R.}\ \bibnamefont
  {{Priest}}},\ }\bibfield  {title} {\enquote {\bibinfo {title} {{Separators in
  3D Quiet-Sun Magnetic Fields}},}\ }\href {\doibase 10.1007/s11207-004-3259-0}
  {\bibfield  {journal} {\bibinfo  {journal} {Sol. Phys.}\ }\textbf {\bibinfo
  {volume} {225}},\ \bibinfo {pages} {21--46} (\bibinfo {year}
  {2004})}\BibitemShut {NoStop}%
\bibitem [{\citenamefont {{R{\'e}gnier}}, \citenamefont {{Parnell}},\ and\
  \citenamefont {{Haynes}}(2008)}]{Regnier:2008}%
  \BibitemOpen
  \bibfield  {author} {\bibinfo {author} {\bibfnamefont {S.}~\bibnamefont
  {{R{\'e}gnier}}}, \bibinfo {author} {\bibfnamefont {C.~E.}\ \bibnamefont
  {{Parnell}}}, \ and\ \bibinfo {author} {\bibfnamefont {A.~L.}\ \bibnamefont
  {{Haynes}}},\ }\bibfield  {title} {\enquote {\bibinfo {title} {{A new view of
  quiet-Sun topology from Hinode/SOT}},}\ }\href {\doibase
  10.1051/0004-6361:200809826} {\bibfield  {journal} {\bibinfo  {journal}
  {Astron. Astrophys.}\ }\textbf {\bibinfo {volume} {484}},\ \bibinfo {pages}
  {L47--L50} (\bibinfo {year} {2008})}\BibitemShut {NoStop}%
\bibitem [{\citenamefont {Servidio}\ \emph {et~al.}(2009)\citenamefont
  {Servidio}, \citenamefont {Matthaeus}, \citenamefont {Shay}, \citenamefont
  {Cassak},\ and\ \citenamefont {Dmitruk}}]{servidio:2009}%
  \BibitemOpen
  \bibfield  {author} {\bibinfo {author} {\bibfnamefont {S.}~\bibnamefont
  {Servidio}}, \bibinfo {author} {\bibfnamefont {W.~H.}\ \bibnamefont
  {Matthaeus}}, \bibinfo {author} {\bibfnamefont {M.~A.}\ \bibnamefont {Shay}},
  \bibinfo {author} {\bibfnamefont {P.~A.}\ \bibnamefont {Cassak}}, \ and\
  \bibinfo {author} {\bibfnamefont {P.}~\bibnamefont {Dmitruk}},\ }\bibfield
  {title} {\enquote {\bibinfo {title} {{Magnetic Reconnection in
  Two-Dimensional Magnetohydrodynamic Turbulence}},}\ }\href {\doibase
  10.1103/PhysRevLett.102.115003} {\bibfield  {journal} {\bibinfo  {journal}
  {Phys. Rev. Lett.}\ }\textbf {\bibinfo {volume} {102}},\ \bibinfo {pages}
  {115003} (\bibinfo {year} {2009})}\BibitemShut {NoStop}%
\bibitem [{\citenamefont {Servidio}\ \emph {et~al.}(2010)\citenamefont
  {Servidio}, \citenamefont {Matthaeus}, \citenamefont {Shay}, \citenamefont
  {Dmitruk}, \citenamefont {Cassak},\ and\ \citenamefont
  {Wan}}]{servidio:2010}%
  \BibitemOpen
  \bibfield  {author} {\bibinfo {author} {\bibfnamefont {S.}~\bibnamefont
  {Servidio}}, \bibinfo {author} {\bibfnamefont {W.~H.}\ \bibnamefont
  {Matthaeus}}, \bibinfo {author} {\bibfnamefont {M.~A.}\ \bibnamefont {Shay}},
  \bibinfo {author} {\bibfnamefont {P.}~\bibnamefont {Dmitruk}}, \bibinfo
  {author} {\bibfnamefont {P.~A.}\ \bibnamefont {Cassak}}, \ and\ \bibinfo
  {author} {\bibfnamefont {M.}~\bibnamefont {Wan}},\ }\bibfield  {title}
  {\enquote {\bibinfo {title} {{Statistics of magnetic reconnection in
  two-dimensional magnetohydrodynamic turbulence}},}\ }\href {\doibase
  10.1063/1.3368798} {\bibfield  {journal} {\bibinfo  {journal} {Phys.\
  Plasmas}\ }\textbf {\bibinfo {volume} {17}},\ \bibinfo {pages} {032315}
  (\bibinfo {year} {2010})}\BibitemShut {NoStop}%
\bibitem [{\citenamefont {{Maclean}}, \citenamefont {{Parnell}},\ and\
  \citenamefont {{Galsgaard}}(2009)}]{Maclean:2009}%
  \BibitemOpen
  \bibfield  {author} {\bibinfo {author} {\bibfnamefont {R.~C.}\ \bibnamefont
  {{Maclean}}}, \bibinfo {author} {\bibfnamefont {C.~E.}\ \bibnamefont
  {{Parnell}}}, \ and\ \bibinfo {author} {\bibfnamefont {K.}~\bibnamefont
  {{Galsgaard}}},\ }\bibfield  {title} {\enquote {\bibinfo {title} {{Is
  Null-Point Reconnection Important for Solar Flux Emergence?}}}\ }\href
  {\doibase 10.1007/s11207-009-9458-y} {\bibfield  {journal} {\bibinfo
  {journal} {Sol. Phys.}\ }\textbf {\bibinfo {volume} {260}},\ \bibinfo {pages}
  {299--320} (\bibinfo {year} {2009})}\BibitemShut {NoStop}%
\bibitem [{\citenamefont {Parnell}, \citenamefont {Maclean},\ and\
  \citenamefont {Haynes}(2010)}]{Parnell:2010B}%
  \BibitemOpen
  \bibfield  {author} {\bibinfo {author} {\bibfnamefont {C.~E.}\ \bibnamefont
  {Parnell}}, \bibinfo {author} {\bibfnamefont {R.~C.}\ \bibnamefont
  {Maclean}}, \ and\ \bibinfo {author} {\bibfnamefont {A.~L.}\ \bibnamefont
  {Haynes}},\ }\bibfield  {title} {\enquote {\bibinfo {title} {{The Detection
  of Numerous Magnetic Separators in a Three-Dimensional Magnetohydrodynamic
  Model of Solar Emerging Flux}},}\ }\href {\doibase
  10.1088/2041-8205/725/2/L214} {\bibfield  {journal} {\bibinfo  {journal}
  {Astrophys. J.}\ }\textbf {\bibinfo {volume} {725}},\ \bibinfo {pages}
  {L214--L218} (\bibinfo {year} {2010})}\BibitemShut {NoStop}%
\bibitem [{\citenamefont {Priest}, \citenamefont {Hornig},\ and\ \citenamefont
  {Pontin}(2003)}]{priest:2003}%
  \BibitemOpen
  \bibfield  {author} {\bibinfo {author} {\bibfnamefont {E.~R.}\ \bibnamefont
  {Priest}}, \bibinfo {author} {\bibfnamefont {G.}~\bibnamefont {Hornig}}, \
  and\ \bibinfo {author} {\bibfnamefont {D.~I.}\ \bibnamefont {Pontin}},\
  }\bibfield  {title} {\enquote {\bibinfo {title} {{On the nature of
  three-dimensional magnetic reconnection}},}\ }\href {\doibase
  10.1029/2002JA009812} {\bibfield  {journal} {\bibinfo  {journal} {J.\
  Geophys.\ Res.}\ }\textbf {\bibinfo {volume} {108}},\ \bibinfo {pages} {1285}
  (\bibinfo {year} {2003})}\BibitemShut {NoStop}%
\bibitem [{\citenamefont {Pontin}(2011)}]{pontin:2011:review}%
  \BibitemOpen
  \bibfield  {author} {\bibinfo {author} {\bibfnamefont {D.~I.}\ \bibnamefont
  {Pontin}},\ }\bibfield  {title} {\enquote {\bibinfo {title}
  {{Three-dimensional magnetic reconnection regimes: A review}},}\ }\href
  {\doibase 10.1016/j.asr.2010.12.022} {\bibfield  {journal} {\bibinfo
  {journal} {Adv.\ Space Res.}\ }\textbf {\bibinfo {volume} {47}},\ \bibinfo
  {pages} {1508--1522} (\bibinfo {year} {2011})}\BibitemShut {NoStop}%
\bibitem [{\citenamefont {Hornig}\ and\ \citenamefont
  {Schindler}(1996)}]{hornig:1996}%
  \BibitemOpen
  \bibfield  {author} {\bibinfo {author} {\bibfnamefont {G.}~\bibnamefont
  {Hornig}}\ and\ \bibinfo {author} {\bibfnamefont {K.}~\bibnamefont
  {Schindler}},\ }\bibfield  {title} {\enquote {\bibinfo {title} {{Magnetic
  topology and the problem of its invariant definition}},}\ }\href {\doibase
  10.1063/1.871778} {\bibfield  {journal} {\bibinfo  {journal} {Phys.\
  Plasmas}\ }\textbf {\bibinfo {volume} {3}},\ \bibinfo {pages} {781--791}
  (\bibinfo {year} {1996})}\BibitemShut {NoStop}%
\bibitem [{\citenamefont {Longcope}(2005)}]{longcope:2005}%
  \BibitemOpen
  \bibfield  {author} {\bibinfo {author} {\bibfnamefont {D.~W.}\ \bibnamefont
  {Longcope}},\ }\bibfield  {title} {\enquote {\bibinfo {title} {{Topological
  Methods for the Analysis of Solar Magnetic Fields}},}\ }\href@noop {}
  {\bibfield  {journal} {\bibinfo  {journal} {LRSP}\ }\textbf {\bibinfo
  {volume} {2}},\ \bibinfo {pages} {7} (\bibinfo {year} {2005})}\BibitemShut
  {NoStop}%
\bibitem [{\citenamefont {Haynes}\ and\ \citenamefont
  {Parnell}(2010)}]{haynes:2010}%
  \BibitemOpen
  \bibfield  {author} {\bibinfo {author} {\bibfnamefont {A.~L.}\ \bibnamefont
  {Haynes}}\ and\ \bibinfo {author} {\bibfnamefont {C.~E.}\ \bibnamefont
  {Parnell}},\ }\bibfield  {title} {\enquote {\bibinfo {title} {{A method for
  finding three-dimensional magnetic skeletons}},}\ }\href {\doibase
  10.1063/1.3467499} {\bibfield  {journal} {\bibinfo  {journal} {Phys.\
  Plasmas}\ }\textbf {\bibinfo {volume} {17}},\ \bibinfo {pages} {092903}
  (\bibinfo {year} {2010})}\BibitemShut {NoStop}%
\bibitem [{\citenamefont {Parnell}, \citenamefont {Haynes},\ and\ \citenamefont
  {Galsgaard}(2010)}]{Parnell:2010A}%
  \BibitemOpen
  \bibfield  {author} {\bibinfo {author} {\bibfnamefont {C.~E.}\ \bibnamefont
  {Parnell}}, \bibinfo {author} {\bibfnamefont {A.~L.}\ \bibnamefont {Haynes}},
  \ and\ \bibinfo {author} {\bibfnamefont {K.}~\bibnamefont {Galsgaard}},\
  }\bibfield  {title} {\enquote {\bibinfo {title} {{Structure of magnetic
  separators and separator reconnection}},}\ }\href {\doibase
  10.1029/2009JA014557} {\bibfield  {journal} {\bibinfo  {journal} {J.\
  Geophys.\ Res.}\ }\textbf {\bibinfo {volume} {115}},\ \bibinfo {pages} {2102}
  (\bibinfo {year} {2010})}\BibitemShut {NoStop}%
\bibitem [{\citenamefont {Hesse}\ and\ \citenamefont
  {Schindler}(1988)}]{Hesse:1988}%
  \BibitemOpen
  \bibfield  {author} {\bibinfo {author} {\bibfnamefont {M.}~\bibnamefont
  {Hesse}}\ and\ \bibinfo {author} {\bibfnamefont {K.}~\bibnamefont
  {Schindler}},\ }\bibfield  {title} {\enquote {\bibinfo {title} {{A
  theoretical foundation of general magnetic reconnection}},}\ }\href {\doibase
  10.1029/JA093iA06p05559} {\bibfield  {journal} {\bibinfo  {journal} {J.\
  Geophys.\ Res.}\ }\textbf {\bibinfo {volume} {93}},\ \bibinfo {pages}
  {5559--5567} (\bibinfo {year} {1988})}\BibitemShut {NoStop}%
\bibitem [{\citenamefont {Schindler}, \citenamefont {Hesse},\ and\
  \citenamefont {Birn}(1988)}]{Schindler:1988}%
  \BibitemOpen
  \bibfield  {author} {\bibinfo {author} {\bibfnamefont {K.}~\bibnamefont
  {Schindler}}, \bibinfo {author} {\bibfnamefont {M.}~\bibnamefont {Hesse}}, \
  and\ \bibinfo {author} {\bibfnamefont {J.}~\bibnamefont {Birn}},\ }\bibfield
  {title} {\enquote {\bibinfo {title} {{General magnetic reconnection, parallel
  electric fields, and helicity}},}\ }\href {\doibase 10.1029/JA093iA06p05547}
  {\bibfield  {journal} {\bibinfo  {journal} {J.\ Geophys.\ Res.}\ }\textbf
  {\bibinfo {volume} {93}},\ \bibinfo {pages} {5547--5557} (\bibinfo {year}
  {1988})}\BibitemShut {NoStop}%
\bibitem [{\citenamefont {Priest}\ and\ \citenamefont
  {Forbes}(1992)}]{priest:1992B}%
  \BibitemOpen
  \bibfield  {author} {\bibinfo {author} {\bibfnamefont {E.~R.}\ \bibnamefont
  {Priest}}\ and\ \bibinfo {author} {\bibfnamefont {T.~G.}\ \bibnamefont
  {Forbes}},\ }\bibfield  {title} {\enquote {\bibinfo {title} {{Magnetic
  flipping - Reconnection in three dimensions without null points}},}\ }\href
  {\doibase 10.1029/91JA02435} {\bibfield  {journal} {\bibinfo  {journal} {J.\
  Geophys.\ Res.}\ }\textbf {\bibinfo {volume} {97}},\ \bibinfo {pages}
  {1521--1531} (\bibinfo {year} {1992})}\BibitemShut {NoStop}%
\bibitem [{\citenamefont {Aulanier}\ \emph {et~al.}(2006)\citenamefont
  {Aulanier}, \citenamefont {Pariat}, \citenamefont {D{\'e}moulin},\ and\
  \citenamefont {DeVore}}]{aulanier:2006}%
  \BibitemOpen
  \bibfield  {author} {\bibinfo {author} {\bibfnamefont {G.}~\bibnamefont
  {Aulanier}}, \bibinfo {author} {\bibfnamefont {E.}~\bibnamefont {Pariat}},
  \bibinfo {author} {\bibfnamefont {P.}~\bibnamefont {D{\'e}moulin}}, \ and\
  \bibinfo {author} {\bibfnamefont {C.~R.}\ \bibnamefont {DeVore}},\ }\bibfield
   {title} {\enquote {\bibinfo {title} {{Slip-Running Reconnection in
  Quasi-Separatrix Layers}},}\ }\href {\doibase 10.1007/s11207-006-0230-2}
  {\bibfield  {journal} {\bibinfo  {journal} {Sol. Phys.}\ }\textbf {\bibinfo
  {volume} {238}},\ \bibinfo {pages} {347--376} (\bibinfo {year}
  {2006})}\BibitemShut {NoStop}%
\bibitem [{\citenamefont {{Janvier}}\ \emph {et~al.}(2013)\citenamefont
  {{Janvier}}, \citenamefont {{Aulanier}}, \citenamefont {{Pariat}},\ and\
  \citenamefont {{D{\'e}moulin}}}]{janvier:2013:III}%
  \BibitemOpen
  \bibfield  {author} {\bibinfo {author} {\bibfnamefont {M.}~\bibnamefont
  {{Janvier}}}, \bibinfo {author} {\bibfnamefont {G.}~\bibnamefont
  {{Aulanier}}}, \bibinfo {author} {\bibfnamefont {E.}~\bibnamefont
  {{Pariat}}}, \ and\ \bibinfo {author} {\bibfnamefont {P.}~\bibnamefont
  {{D{\'e}moulin}}},\ }\bibfield  {title} {\enquote {\bibinfo {title} {{The
  standard flare model in three dimensions. III. Slip-running reconnection
  properties}},}\ }\href {\doibase 10.1051/0004-6361/201321164} {\bibfield
  {journal} {\bibinfo  {journal} {Astron. Astrophys.}\ }\textbf {\bibinfo
  {volume} {555}},\ \bibinfo {eid} {A77} (\bibinfo {year} {2013})}\BibitemShut
  {NoStop}%
\bibitem [{\citenamefont {Forbes}\ \emph {et~al.}(1981)\citenamefont {Forbes},
  \citenamefont {Hones}, \citenamefont {Bame}, \citenamefont {Asbridge},
  \citenamefont {Paschmann}, \citenamefont {Sckopke},\ and\ \citenamefont
  {Russell}}]{forbes:1981}%
  \BibitemOpen
  \bibfield  {author} {\bibinfo {author} {\bibfnamefont {T.~G.}\ \bibnamefont
  {Forbes}}, \bibinfo {author} {\bibfnamefont {E.~W.}\ \bibnamefont {Hones}},
  \bibinfo {author} {\bibfnamefont {S.~J.}\ \bibnamefont {Bame}}, \bibinfo
  {author} {\bibfnamefont {J.~R.}\ \bibnamefont {Asbridge}}, \bibinfo {author}
  {\bibfnamefont {G.}~\bibnamefont {Paschmann}}, \bibinfo {author}
  {\bibfnamefont {N.}~\bibnamefont {Sckopke}}, \ and\ \bibinfo {author}
  {\bibfnamefont {C.~T.}\ \bibnamefont {Russell}},\ }\bibfield  {title}
  {\enquote {\bibinfo {title} {{Evidence for the tailward retreat of a magnetic
  neutral line in the magnetotail during substorm recovery}},}\ }\href
  {\doibase 10.1029/GL008i003p00261} {\bibfield  {journal} {\bibinfo  {journal}
  {Geophys. Res. Lett.}\ }\textbf {\bibinfo {volume} {8}},\ \bibinfo {pages}
  {261--264} (\bibinfo {year} {1981})}\BibitemShut {NoStop}%
\bibitem [{\citenamefont {{Hasegawa}}\ \emph {et~al.}(2008)\citenamefont
  {{Hasegawa}}, \citenamefont {{Retin{\`o}}}, \citenamefont {{Vaivads}},
  \citenamefont {{Khotyaintsev}}, \citenamefont {{Nakamura}}, \citenamefont
  {{Takada}}, \citenamefont {{Miyashita}}, \citenamefont {{R{\`e}me}},\ and\
  \citenamefont {{Lucek}}}]{hasegawa:2008}%
  \BibitemOpen
  \bibfield  {author} {\bibinfo {author} {\bibfnamefont {H.}~\bibnamefont
  {{Hasegawa}}}, \bibinfo {author} {\bibfnamefont {A.}~\bibnamefont
  {{Retin{\`o}}}}, \bibinfo {author} {\bibfnamefont {A.}~\bibnamefont
  {{Vaivads}}}, \bibinfo {author} {\bibfnamefont {Y.}~\bibnamefont
  {{Khotyaintsev}}}, \bibinfo {author} {\bibfnamefont {R.}~\bibnamefont
  {{Nakamura}}}, \bibinfo {author} {\bibfnamefont {T.}~\bibnamefont
  {{Takada}}}, \bibinfo {author} {\bibfnamefont {Y.}~\bibnamefont
  {{Miyashita}}}, \bibinfo {author} {\bibfnamefont {H.}~\bibnamefont
  {{R{\`e}me}}}, \ and\ \bibinfo {author} {\bibfnamefont {E.~A.}\ \bibnamefont
  {{Lucek}}},\ }\bibfield  {title} {\enquote {\bibinfo {title} {{Retreat and
  reformation of X-line during quasi-continuous tailward-of-the-cusp
  reconnection under northward IMF}},}\ }\href {\doibase 10.1029/2008GL034767}
  {\bibfield  {journal} {\bibinfo  {journal} {Geophys. Res. Lett.}\ }\textbf
  {\bibinfo {volume} {35}},\ \bibinfo {eid} {L15104} (\bibinfo {year}
  {2008})}\BibitemShut {NoStop}%
\bibitem [{\citenamefont {Oka}\ \emph {et~al.}(2011)\citenamefont {Oka},
  \citenamefont {Phan}, \citenamefont {Eastwood}, \citenamefont {Angelopoulos},
  \citenamefont {Murphy}, \citenamefont {{\O}ieroset}, \citenamefont
  {Miyashita}, \citenamefont {Fujimoto}, \citenamefont {{McFadden}},\ and\
  \citenamefont {Larson}}]{oka:2011}%
  \BibitemOpen
  \bibfield  {author} {\bibinfo {author} {\bibfnamefont {M.}~\bibnamefont
  {Oka}}, \bibinfo {author} {\bibfnamefont {T.-D.}\ \bibnamefont {Phan}},
  \bibinfo {author} {\bibfnamefont {J.~P.}\ \bibnamefont {Eastwood}}, \bibinfo
  {author} {\bibfnamefont {V.}~\bibnamefont {Angelopoulos}}, \bibinfo {author}
  {\bibfnamefont {N.~A.}\ \bibnamefont {Murphy}}, \bibinfo {author}
  {\bibfnamefont {M.}~\bibnamefont {{\O}ieroset}}, \bibinfo {author}
  {\bibfnamefont {Y.}~\bibnamefont {Miyashita}}, \bibinfo {author}
  {\bibfnamefont {M.}~\bibnamefont {Fujimoto}}, \bibinfo {author}
  {\bibfnamefont {J.}~\bibnamefont {{McFadden}}}, \ and\ \bibinfo {author}
  {\bibfnamefont {D.}~\bibnamefont {Larson}},\ }\bibfield  {title} {\enquote
  {\bibinfo {title} {{Magnetic reconnection X-line retreat associated with
  dipolarization of the Earth's magnetosphere}},}\ }\href {\doibase
  10.1029/2011GL049350} {\bibfield  {journal} {\bibinfo  {journal} {Geophys.
  Res. Lett.}\ }\textbf {\bibinfo {volume} {38}},\ \bibinfo {pages} {20105}
  (\bibinfo {year} {2011})}\BibitemShut {NoStop}%
\bibitem [{\citenamefont {{Cao}}\ \emph {et~al.}(2012)\citenamefont {{Cao}},
  \citenamefont {{Pu}}, \citenamefont {{Du}}, \citenamefont {{Mishin}},
  \citenamefont {{Wang}}, \citenamefont {{Xiao}}, \citenamefont {{Zhang}},
  \citenamefont {{Angelopoulos}}, \citenamefont {{McFadden}},\ and\
  \citenamefont {{Glassmeier}}}]{xcao:2012}%
  \BibitemOpen
  \bibfield  {author} {\bibinfo {author} {\bibfnamefont {X.}~\bibnamefont
  {{Cao}}}, \bibinfo {author} {\bibfnamefont {Z.~Y.}\ \bibnamefont {{Pu}}},
  \bibinfo {author} {\bibfnamefont {A.~M.}\ \bibnamefont {{Du}}}, \bibinfo
  {author} {\bibfnamefont {V.~M.}\ \bibnamefont {{Mishin}}}, \bibinfo {author}
  {\bibfnamefont {X.~G.}\ \bibnamefont {{Wang}}}, \bibinfo {author}
  {\bibfnamefont {C.~J.}\ \bibnamefont {{Xiao}}}, \bibinfo {author}
  {\bibfnamefont {T.~L.}\ \bibnamefont {{Zhang}}}, \bibinfo {author}
  {\bibfnamefont {V.}~\bibnamefont {{Angelopoulos}}}, \bibinfo {author}
  {\bibfnamefont {J.~P.}\ \bibnamefont {{McFadden}}}, \ and\ \bibinfo {author}
  {\bibfnamefont {K.~H.}\ \bibnamefont {{Glassmeier}}},\ }\bibfield  {title}
  {\enquote {\bibinfo {title} {{On the retreat of near-Earth neutral line
  during substorm expansion phase: a THEMIS case study during the 9 January
  2008 substorm}},}\ }\href {\doibase 10.5194/angeo-30-143-2012} {\bibfield
  {journal} {\bibinfo  {journal} {Ann. Geophys.}\ }\textbf {\bibinfo {volume}
  {30}},\ \bibinfo {pages} {143--151} (\bibinfo {year} {2012})}\BibitemShut
  {NoStop}%
\bibitem [{\citenamefont {{Wilder}}\ \emph {et~al.}(2014)\citenamefont
  {{Wilder}}, \citenamefont {{Eriksson}}, \citenamefont {{Trattner}},
  \citenamefont {{Cassak}}, \citenamefont {{Fuselier}},\ and\ \citenamefont
  {{Lybekk}}}]{wilder:2014}%
  \BibitemOpen
  \bibfield  {author} {\bibinfo {author} {\bibfnamefont {F.~D.}\ \bibnamefont
  {{Wilder}}}, \bibinfo {author} {\bibfnamefont {S.}~\bibnamefont
  {{Eriksson}}}, \bibinfo {author} {\bibfnamefont {K.~J.}\ \bibnamefont
  {{Trattner}}}, \bibinfo {author} {\bibfnamefont {P.~A.}\ \bibnamefont
  {{Cassak}}}, \bibinfo {author} {\bibfnamefont {S.~A.}\ \bibnamefont
  {{Fuselier}}}, \ and\ \bibinfo {author} {\bibfnamefont {B.}~\bibnamefont
  {{Lybekk}}},\ }\bibfield  {title} {\enquote {\bibinfo {title} {{Observation
  of a retreating x line and magnetic islands poleward of the cusp during
  northward interplanetary magnetic field conditions}},}\ }\href {\doibase
  10.1002/2014JA020453} {\bibfield  {journal} {\bibinfo  {journal} {J.\
  Geophys.\ Res.}\ }\textbf {\bibinfo {volume} {119}},\ \bibinfo {pages}
  {9643--9657} (\bibinfo {year} {2014})}\BibitemShut {NoStop}%
\bibitem [{\citenamefont {Swisdak}\ \emph {et~al.}(2003)\citenamefont
  {Swisdak}, \citenamefont {Rogers}, \citenamefont {Drake},\ and\ \citenamefont
  {Shay}}]{swisdak:2003}%
  \BibitemOpen
  \bibfield  {author} {\bibinfo {author} {\bibfnamefont {M.}~\bibnamefont
  {Swisdak}}, \bibinfo {author} {\bibfnamefont {B.~N.}\ \bibnamefont {Rogers}},
  \bibinfo {author} {\bibfnamefont {J.~F.}\ \bibnamefont {Drake}}, \ and\
  \bibinfo {author} {\bibfnamefont {M.~A.}\ \bibnamefont {Shay}},\ }\bibfield
  {title} {\enquote {\bibinfo {title} {{Diamagnetic suppression of component
  magnetic reconnection at the magnetopause}},}\ }\href {\doibase
  10.1029/2002JA009726} {\bibfield  {journal} {\bibinfo  {journal} {J.\
  Geophys.\ Res.}\ }\textbf {\bibinfo {volume} {108}},\ \bibinfo {pages} {1218}
  (\bibinfo {year} {2003})}\BibitemShut {NoStop}%
\bibitem [{\citenamefont {{Phan}}\ \emph {et~al.}(2013)\citenamefont {{Phan}},
  \citenamefont {{Paschmann}}, \citenamefont {{Gosling}}, \citenamefont
  {{Oieroset}}, \citenamefont {{Fujimoto}}, \citenamefont {{Drake}},\ and\
  \citenamefont {{Angelopoulos}}}]{phan:2013}%
  \BibitemOpen
  \bibfield  {author} {\bibinfo {author} {\bibfnamefont {T.~D.}\ \bibnamefont
  {{Phan}}}, \bibinfo {author} {\bibfnamefont {G.}~\bibnamefont {{Paschmann}}},
  \bibinfo {author} {\bibfnamefont {J.~T.}\ \bibnamefont {{Gosling}}}, \bibinfo
  {author} {\bibfnamefont {M.}~\bibnamefont {{Oieroset}}}, \bibinfo {author}
  {\bibfnamefont {M.}~\bibnamefont {{Fujimoto}}}, \bibinfo {author}
  {\bibfnamefont {J.~F.}\ \bibnamefont {{Drake}}}, \ and\ \bibinfo {author}
  {\bibfnamefont {V.}~\bibnamefont {{Angelopoulos}}},\ }\bibfield  {title}
  {\enquote {\bibinfo {title} {{The dependence of magnetic reconnection on
  plasma {$\beta$} and magnetic shear: Evidence from magnetopause
  observations}},}\ }\href {\doibase 10.1029/2012GL054528} {\bibfield
  {journal} {\bibinfo  {journal} {Geophys. Res. Lett.}\ }\textbf {\bibinfo
  {volume} {40}},\ \bibinfo {pages} {11--16} (\bibinfo {year}
  {2013})}\BibitemShut {NoStop}%
\bibitem [{\citenamefont {Rogers}\ and\ \citenamefont
  {Zakharov}(1995)}]{rogers:1995}%
  \BibitemOpen
  \bibfield  {author} {\bibinfo {author} {\bibfnamefont {B.}~\bibnamefont
  {Rogers}}\ and\ \bibinfo {author} {\bibfnamefont {L.}~\bibnamefont
  {Zakharov}},\ }\bibfield  {title} {\enquote {\bibinfo {title} {{Nonlinear
  $\omega_*$-stabilization of the $m${$=$}$1$ mode in tokamaks}},}\ }\href
  {\doibase 10.1063/1.871124} {\bibfield  {journal} {\bibinfo  {journal}
  {Phys.\ Plasmas}\ }\textbf {\bibinfo {volume} {2}},\ \bibinfo {pages}
  {3420--3428} (\bibinfo {year} {1995})}\BibitemShut {NoStop}%
\bibitem [{\citenamefont {Beidler}\ and\ \citenamefont
  {Cassak}(2011)}]{beidler:2011}%
  \BibitemOpen
  \bibfield  {author} {\bibinfo {author} {\bibfnamefont {M.~T.}\ \bibnamefont
  {Beidler}}\ and\ \bibinfo {author} {\bibfnamefont {P.~A.}\ \bibnamefont
  {Cassak}},\ }\bibfield  {title} {\enquote {\bibinfo {title} {{Model for
  Incomplete Reconnection in Sawtooth Crashes}},}\ }\href {\doibase
  10.1103/PhysRevLett.107.255002} {\bibfield  {journal} {\bibinfo  {journal}
  {Phys. Rev. Lett.}\ }\textbf {\bibinfo {volume} {107}},\ \bibinfo {pages}
  {255002} (\bibinfo {year} {2011})}\BibitemShut {NoStop}%
\bibitem [{\citenamefont {Inomoto}\ \emph {et~al.}(2006)\citenamefont
  {Inomoto}, \citenamefont {Gerhardt}, \citenamefont {Yamada}, \citenamefont
  {Ji}, \citenamefont {Belova}, \citenamefont {Kuritsyn},\ and\ \citenamefont
  {Ren}}]{inomoto:counter}%
  \BibitemOpen
  \bibfield  {author} {\bibinfo {author} {\bibfnamefont {M.}~\bibnamefont
  {Inomoto}}, \bibinfo {author} {\bibfnamefont {S.~P.}\ \bibnamefont
  {Gerhardt}}, \bibinfo {author} {\bibfnamefont {M.}~\bibnamefont {Yamada}},
  \bibinfo {author} {\bibfnamefont {H.}~\bibnamefont {Ji}}, \bibinfo {author}
  {\bibfnamefont {E.}~\bibnamefont {Belova}}, \bibinfo {author} {\bibfnamefont
  {A.}~\bibnamefont {Kuritsyn}}, \ and\ \bibinfo {author} {\bibfnamefont
  {Y.}~\bibnamefont {Ren}},\ }\bibfield  {title} {\enquote {\bibinfo {title}
  {{Coupling between Global Geometry and the Local Hall Effect Leading to
  Reconnection-Layer Symmetry Breaking}},}\ }\href {\doibase
  10.1103/PhysRevLett.97.135002} {\bibfield  {journal} {\bibinfo  {journal}
  {Phys. Rev. Lett.}\ }\textbf {\bibinfo {volume} {97}},\ \bibinfo {pages}
  {135002} (\bibinfo {year} {2006})}\BibitemShut {NoStop}%
\bibitem [{\citenamefont {{Yoo}}\ \emph {et~al.}(2014)\citenamefont {{Yoo}},
  \citenamefont {{Yamada}}, \citenamefont {{Ji}}, \citenamefont
  {{Jara-Almonte}}, \citenamefont {{Myers}},\ and\ \citenamefont
  {{Chen}}}]{yoo:2014}%
  \BibitemOpen
  \bibfield  {author} {\bibinfo {author} {\bibfnamefont {J.}~\bibnamefont
  {{Yoo}}}, \bibinfo {author} {\bibfnamefont {M.}~\bibnamefont {{Yamada}}},
  \bibinfo {author} {\bibfnamefont {H.}~\bibnamefont {{Ji}}}, \bibinfo {author}
  {\bibfnamefont {J.}~\bibnamefont {{Jara-Almonte}}}, \bibinfo {author}
  {\bibfnamefont {C.~E.}\ \bibnamefont {{Myers}}}, \ and\ \bibinfo {author}
  {\bibfnamefont {L.-J.}\ \bibnamefont {{Chen}}},\ }\bibfield  {title}
  {\enquote {\bibinfo {title} {{Laboratory Study of Magnetic Reconnection with
  a Density Asymmetry across the Current Sheet}},}\ }\href {\doibase
  10.1103/PhysRevLett.113.095002} {\bibfield  {journal} {\bibinfo  {journal}
  {Phys. Rev. Lett.}\ }\textbf {\bibinfo {volume} {113}},\ \bibinfo {eid}
  {095002} (\bibinfo {year} {2014})}\BibitemShut {NoStop}%
\bibitem [{\citenamefont {Murphy}\ and\ \citenamefont
  {Sovinec}(2008)}]{murphy:mrx}%
  \BibitemOpen
  \bibfield  {author} {\bibinfo {author} {\bibfnamefont {N.~A.}\ \bibnamefont
  {Murphy}}\ and\ \bibinfo {author} {\bibfnamefont {C.~R.}\ \bibnamefont
  {Sovinec}},\ }\bibfield  {title} {\enquote {\bibinfo {title} {{Global
  axisymmetric simulations of two-fluid reconnection in an experimentally
  relevant geometry}},}\ }\href {\doibase 10.1063/1.2904600} {\bibfield
  {journal} {\bibinfo  {journal} {Phys.\ Plasmas}\ }\textbf {\bibinfo {volume}
  {15}},\ \bibinfo {pages} {042313} (\bibinfo {year} {2008})}\BibitemShut
  {NoStop}%
\bibitem [{\citenamefont {Lukin}\ and\ \citenamefont
  {Linton}(2011)}]{lukin:2011}%
  \BibitemOpen
  \bibfield  {author} {\bibinfo {author} {\bibfnamefont {V.~S.}\ \bibnamefont
  {Lukin}}\ and\ \bibinfo {author} {\bibfnamefont {M.~G.}\ \bibnamefont
  {Linton}},\ }\bibfield  {title} {\enquote {\bibinfo {title}
  {{Three-dimensional magnetic reconnection through a moving magnetic null}},}\
  }\href {\doibase 10.5194/npg-18-871-2011} {\bibfield  {journal} {\bibinfo
  {journal} {NPGeo}\ }\textbf {\bibinfo {volume} {18}},\ \bibinfo {pages}
  {871--882} (\bibinfo {year} {2011})}\BibitemShut {NoStop}%
\bibitem [{\citenamefont {Forbes}\ and\ \citenamefont
  {Acton}(1996)}]{forbes:1996}%
  \BibitemOpen
  \bibfield  {author} {\bibinfo {author} {\bibfnamefont {T.~G.}\ \bibnamefont
  {Forbes}}\ and\ \bibinfo {author} {\bibfnamefont {L.~W.}\ \bibnamefont
  {Acton}},\ }\bibfield  {title} {\enquote {\bibinfo {title} {{Reconnection and
  Field Line Shrinkage in Solar Flares}},}\ }\href {\doibase 10.1086/176896}
  {\bibfield  {journal} {\bibinfo  {journal} {Astrophys. J.}\ }\textbf
  {\bibinfo {volume} {459}},\ \bibinfo {pages} {330} (\bibinfo {year}
  {1996})}\BibitemShut {NoStop}%
\bibitem [{\citenamefont {Savage}\ \emph {et~al.}(2010)\citenamefont {Savage},
  \citenamefont {{McKenzie}}, \citenamefont {Reeves}, \citenamefont {Forbes},\
  and\ \citenamefont {Longcope}}]{savage:2010}%
  \BibitemOpen
  \bibfield  {author} {\bibinfo {author} {\bibfnamefont {S.~L.}\ \bibnamefont
  {Savage}}, \bibinfo {author} {\bibfnamefont {D.~E.}\ \bibnamefont
  {{McKenzie}}}, \bibinfo {author} {\bibfnamefont {K.~K.}\ \bibnamefont
  {Reeves}}, \bibinfo {author} {\bibfnamefont {T.~G.}\ \bibnamefont {Forbes}},
  \ and\ \bibinfo {author} {\bibfnamefont {D.~W.}\ \bibnamefont {Longcope}},\
  }\bibfield  {title} {\enquote {\bibinfo {title} {{Reconnection Outflows and
  Current Sheet Observed with Hinode/XRT in the 2008 April 9 `Cartwheel CME'
  Flare}},}\ }\href {\doibase 10.1088/0004-637X/722/1/329} {\bibfield
  {journal} {\bibinfo  {journal} {Astrophys. J.}\ }\textbf {\bibinfo {volume}
  {722}},\ \bibinfo {pages} {329--342} (\bibinfo {year} {2010})}\BibitemShut
  {NoStop}%
\bibitem [{\citenamefont {Cassak}\ and\ \citenamefont
  {Shay}(2007)}]{cassak:asym}%
  \BibitemOpen
  \bibfield  {author} {\bibinfo {author} {\bibfnamefont {P.~A.}\ \bibnamefont
  {Cassak}}\ and\ \bibinfo {author} {\bibfnamefont {M.~A.}\ \bibnamefont
  {Shay}},\ }\bibfield  {title} {\enquote {\bibinfo {title} {{Scaling of
  asymmetric magnetic reconnection: General theory and collisional
  simulations}},}\ }\href {\doibase 10.1063/1.2795630} {\bibfield  {journal}
  {\bibinfo  {journal} {Phys.\ Plasmas}\ }\textbf {\bibinfo {volume} {14}},\
  \bibinfo {pages} {102114} (\bibinfo {year} {2007})}\BibitemShut {NoStop}%
\bibitem [{\citenamefont {Cassak}\ and\ \citenamefont
  {Shay}(2008)}]{cassak:hall}%
  \BibitemOpen
  \bibfield  {author} {\bibinfo {author} {\bibfnamefont {P.~A.}\ \bibnamefont
  {Cassak}}\ and\ \bibinfo {author} {\bibfnamefont {M.~A.}\ \bibnamefont
  {Shay}},\ }\bibfield  {title} {\enquote {\bibinfo {title} {{Scaling of
  asymmetric Hall magnetic reconnection}},}\ }\href {\doibase
  10.1029/2008GL035268} {\bibfield  {journal} {\bibinfo  {journal} {Geophys.
  Res. Lett.}\ }\textbf {\bibinfo {volume} {35}},\ \bibinfo {pages} {19102}
  (\bibinfo {year} {2008})}\BibitemShut {NoStop}%
\bibitem [{\citenamefont {Cassak}\ and\ \citenamefont
  {Shay}(2009)}]{cassak:dissipation}%
  \BibitemOpen
  \bibfield  {author} {\bibinfo {author} {\bibfnamefont {P.~A.}\ \bibnamefont
  {Cassak}}\ and\ \bibinfo {author} {\bibfnamefont {M.~A.}\ \bibnamefont
  {Shay}},\ }\bibfield  {title} {\enquote {\bibinfo {title} {{Structure of the
  dissipation region in fluid simulations of asymmetric magnetic
  reconnection}},}\ }\href {\doibase 10.1063/1.3086867} {\bibfield  {journal}
  {\bibinfo  {journal} {Phys.\ Plasmas}\ }\textbf {\bibinfo {volume} {16}},\
  \bibinfo {pages} {055704} (\bibinfo {year} {2009})}\BibitemShut {NoStop}%
\bibitem [{\citenamefont {Murphy}, \citenamefont {Sovinec},\ and\ \citenamefont
  {Cassak}(2010)}]{murphy:asym}%
  \BibitemOpen
  \bibfield  {author} {\bibinfo {author} {\bibfnamefont {N.~A.}\ \bibnamefont
  {Murphy}}, \bibinfo {author} {\bibfnamefont {C.~R.}\ \bibnamefont {Sovinec}},
  \ and\ \bibinfo {author} {\bibfnamefont {P.~A.}\ \bibnamefont {Cassak}},\
  }\bibfield  {title} {\enquote {\bibinfo {title} {{Magnetic reconnection with
  asymmetry in the outflow direction}},}\ }\href {\doibase
  10.1029/2009JA015183} {\bibfield  {journal} {\bibinfo  {journal} {J.\
  Geophys.\ Res.}\ }\textbf {\bibinfo {volume} {115}},\ \bibinfo {pages} {9206}
  (\bibinfo {year} {2010})}\BibitemShut {NoStop}%
\bibitem [{\citenamefont {Murphy}(2010)}]{murphy:retreat}%
  \BibitemOpen
  \bibfield  {author} {\bibinfo {author} {\bibfnamefont {N.~A.}\ \bibnamefont
  {Murphy}},\ }\bibfield  {title} {\enquote {\bibinfo {title} {{Resistive
  magnetohydrodynamic simulations of X-line retreat during magnetic
  reconnection}},}\ }\href {\doibase 10.1063/1.3494570} {\bibfield  {journal}
  {\bibinfo  {journal} {Phys.\ Plasmas}\ }\textbf {\bibinfo {volume} {17}},\
  \bibinfo {pages} {112310} (\bibinfo {year} {2010})}\BibitemShut {NoStop}%
\bibitem [{\citenamefont {Oka}\ \emph {et~al.}(2008)\citenamefont {Oka},
  \citenamefont {Fujimoto}, \citenamefont {Nakamura}, \citenamefont
  {Shinohara},\ and\ \citenamefont {Nishikawa}}]{oka:2008}%
  \BibitemOpen
  \bibfield  {author} {\bibinfo {author} {\bibfnamefont {M.}~\bibnamefont
  {Oka}}, \bibinfo {author} {\bibfnamefont {M.}~\bibnamefont {Fujimoto}},
  \bibinfo {author} {\bibfnamefont {T.~K.~M.}\ \bibnamefont {Nakamura}},
  \bibinfo {author} {\bibfnamefont {I.}~\bibnamefont {Shinohara}}, \ and\
  \bibinfo {author} {\bibfnamefont {K.-I.}\ \bibnamefont {Nishikawa}},\
  }\bibfield  {title} {\enquote {\bibinfo {title} {{Magnetic Reconnection by a
  Self-Retreating X Line}},}\ }\href {\doibase 10.1103/PhysRevLett.101.205004}
  {\bibfield  {journal} {\bibinfo  {journal} {Phys. Rev. Lett.}\ }\textbf
  {\bibinfo {volume} {101}},\ \bibinfo {pages} {205004} (\bibinfo {year}
  {2008})}\BibitemShut {NoStop}%
\bibitem [{\citenamefont {{Murphy}}\ and\ \citenamefont
  {{Lukin}}(2015)}]{murphy:partialasym}%
  \BibitemOpen
  \bibfield  {author} {\bibinfo {author} {\bibfnamefont {N.~A.}\ \bibnamefont
  {{Murphy}}}\ and\ \bibinfo {author} {\bibfnamefont {V.~S.}\ \bibnamefont
  {{Lukin}}},\ }\bibfield  {title} {\enquote {\bibinfo {title} {{Asymmetric
  Magnetic Reconnection in Weakly Ionized Chromospheric Plasmas}},}\ }\href
  {\doibase 10.1088/0004-637X/805/2/134} {\bibfield  {journal} {\bibinfo
  {journal} {Astrophys. J.}\ }\textbf {\bibinfo {volume} {805}},\ \bibinfo
  {eid} {134} (\bibinfo {year} {2015})}\BibitemShut {NoStop}%
\bibitem [{\citenamefont {{Murphy}}\ \emph {et~al.}(2012)\citenamefont
  {{Murphy}}, \citenamefont {{Miralles}}, \citenamefont {{Pope}}, \citenamefont
  {{Raymond}}, \citenamefont {{Winter}}, \citenamefont {{Reeves}},
  \citenamefont {{Seaton}}, \citenamefont {{van Ballegooijen}},\ and\
  \citenamefont {{Lin}}}]{murphy:double}%
  \BibitemOpen
  \bibfield  {author} {\bibinfo {author} {\bibfnamefont {N.~A.}\ \bibnamefont
  {{Murphy}}}, \bibinfo {author} {\bibfnamefont {M.~P.}\ \bibnamefont
  {{Miralles}}}, \bibinfo {author} {\bibfnamefont {C.~L.}\ \bibnamefont
  {{Pope}}}, \bibinfo {author} {\bibfnamefont {J.~C.}\ \bibnamefont
  {{Raymond}}}, \bibinfo {author} {\bibfnamefont {H.~D.}\ \bibnamefont
  {{Winter}}}, \bibinfo {author} {\bibfnamefont {K.~K.}\ \bibnamefont
  {{Reeves}}}, \bibinfo {author} {\bibfnamefont {D.~B.}\ \bibnamefont
  {{Seaton}}}, \bibinfo {author} {\bibfnamefont {A.~A.}\ \bibnamefont {{van
  Ballegooijen}}}, \ and\ \bibinfo {author} {\bibfnamefont {J.}~\bibnamefont
  {{Lin}}},\ }\bibfield  {title} {\enquote {\bibinfo {title} {{Asymmetric
  Magnetic Reconnection in Solar Flare and Coronal Mass Ejection Current
  Sheets}},}\ }\href {\doibase 10.1088/0004-637X/751/1/56} {\bibfield
  {journal} {\bibinfo  {journal} {Astrophys. J.}\ }\textbf {\bibinfo {volume}
  {751}},\ \bibinfo {pages} {56} (\bibinfo {year} {2012})}\BibitemShut
  {NoStop}%
\bibitem [{\citenamefont {Siscoe}\ \emph {et~al.}(2002)\citenamefont {Siscoe},
  \citenamefont {Erickson}, \citenamefont {Sonnerup}, \citenamefont {Maynard},
  \citenamefont {Schoendorf}, \citenamefont {Siebert}, \citenamefont {Weimer},
  \citenamefont {White},\ and\ \citenamefont {Wilson}}]{siscoe:2002}%
  \BibitemOpen
  \bibfield  {author} {\bibinfo {author} {\bibfnamefont {G.~L.}\ \bibnamefont
  {Siscoe}}, \bibinfo {author} {\bibfnamefont {G.~M.}\ \bibnamefont
  {Erickson}}, \bibinfo {author} {\bibfnamefont {B.~U.~{\"O}.}\ \bibnamefont
  {Sonnerup}}, \bibinfo {author} {\bibfnamefont {N.~C.}\ \bibnamefont
  {Maynard}}, \bibinfo {author} {\bibfnamefont {J.~A.}\ \bibnamefont
  {Schoendorf}}, \bibinfo {author} {\bibfnamefont {K.~D.}\ \bibnamefont
  {Siebert}}, \bibinfo {author} {\bibfnamefont {D.~R.}\ \bibnamefont {Weimer}},
  \bibinfo {author} {\bibfnamefont {W.~W.}\ \bibnamefont {White}}, \ and\
  \bibinfo {author} {\bibfnamefont {G.~R.}\ \bibnamefont {Wilson}},\ }\bibfield
   {title} {\enquote {\bibinfo {title} {{Flow-through magnetic
  reconnection}},}\ }\href {\doibase 10.1029/2001GL013536} {\bibfield
  {journal} {\bibinfo  {journal} {Geophys. Res. Lett.}\ }\textbf {\bibinfo
  {volume} {29}},\ \bibinfo {pages} {130000--1} (\bibinfo {year}
  {2002})}\BibitemShut {NoStop}%
\bibitem [{\citenamefont {{Maynard}}\ \emph {et~al.}(2012)\citenamefont
  {{Maynard}}, \citenamefont {{Farrugia}}, \citenamefont {{Burke}},
  \citenamefont {{Ober}}, \citenamefont {{Mozer}}, \citenamefont {{R{\`e}me}},
  \citenamefont {{Dunlop}},\ and\ \citenamefont {{Siebert}}}]{maynard:2012}%
  \BibitemOpen
  \bibfield  {author} {\bibinfo {author} {\bibfnamefont {N.~C.}\ \bibnamefont
  {{Maynard}}}, \bibinfo {author} {\bibfnamefont {C.~J.}\ \bibnamefont
  {{Farrugia}}}, \bibinfo {author} {\bibfnamefont {W.~J.}\ \bibnamefont
  {{Burke}}}, \bibinfo {author} {\bibfnamefont {D.~M.}\ \bibnamefont {{Ober}}},
  \bibinfo {author} {\bibfnamefont {F.~S.}\ \bibnamefont {{Mozer}}}, \bibinfo
  {author} {\bibfnamefont {H.}~\bibnamefont {{R{\`e}me}}}, \bibinfo {author}
  {\bibfnamefont {M.}~\bibnamefont {{Dunlop}}}, \ and\ \bibinfo {author}
  {\bibfnamefont {K.~D.}\ \bibnamefont {{Siebert}}},\ }\bibfield  {title}
  {\enquote {\bibinfo {title} {{Cluster observations of the dusk flank
  magnetopause near the sash: Ion dynamics and flow-through reconnection}},}\
  }\href {\doibase 10.1029/2012JA017703} {\bibfield  {journal} {\bibinfo
  {journal} {J.\ Geophys.\ Res.}\ }\textbf {\bibinfo {volume} {117}},\ \bibinfo
  {eid} {A10201} (\bibinfo {year} {2012})}\BibitemShut {NoStop}%
\bibitem [{\citenamefont {{Greene}}(1993)}]{greene:1993}%
  \BibitemOpen
  \bibfield  {author} {\bibinfo {author} {\bibfnamefont {J.~M.}\ \bibnamefont
  {{Greene}}},\ }\bibfield  {title} {\enquote {\bibinfo {title} {{Reconnection
  of vorticity lines and magnetic lines}},}\ }\href {\doibase 10.1063/1.860718}
  {\bibfield  {journal} {\bibinfo  {journal} {Phys.\ Fluids B}\ }\textbf
  {\bibinfo {volume} {5}},\ \bibinfo {pages} {2355--2362} (\bibinfo {year}
  {1993})}\BibitemShut {NoStop}%
\bibitem [{\citenamefont {Lindeberg}(1994)}]{lindeberg:1994}%
  \BibitemOpen
  \bibfield  {author} {\bibinfo {author} {\bibfnamefont {T.}~\bibnamefont
  {Lindeberg}},\ }\bibfield  {title} {\enquote {\bibinfo {title} {Scale-space
  theory: a basic tool for analyzing structures at different scales},}\ }\href
  {\doibase 10.1080/757582976} {\bibfield  {journal} {\bibinfo  {journal}
  {Journal of Applied Statistics}\ }\textbf {\bibinfo {volume} {21}},\ \bibinfo
  {pages} {225--270} (\bibinfo {year} {1994})}\BibitemShut {NoStop}%
\bibitem [{\citenamefont {Klein}\ and\ \citenamefont
  {Ertl}(2007)}]{klein:2007}%
  \BibitemOpen
  \bibfield  {author} {\bibinfo {author} {\bibfnamefont {T.}~\bibnamefont
  {Klein}}\ and\ \bibinfo {author} {\bibfnamefont {T.}~\bibnamefont {Ertl}},\
  }\bibfield  {title} {\enquote {\bibinfo {title} {Scale-space tracking of
  critical points in 3d vector fields},}\ }in\ \href {\doibase
  10.1007/978-3-540-70823-0_3} {\emph {\bibinfo {booktitle} {Topology-based
  Methods in Visualization}}},\ \bibinfo {series and number} {Mathematics and
  Visualization},\ \bibinfo {editor} {edited by\ \bibinfo {editor}
  {\bibfnamefont {H.}~\bibnamefont {Hauser}}, \bibinfo {editor} {\bibfnamefont
  {H.}~\bibnamefont {Hagen}}, \ and\ \bibinfo {editor} {\bibfnamefont
  {H.}~\bibnamefont {Theisel}}}\ (\bibinfo  {publisher} {Springer Berlin
  Heidelberg},\ \bibinfo {year} {2007})\ pp.\ \bibinfo {pages}
  {35--49}\BibitemShut {NoStop}%
\bibitem [{\citenamefont {Priest}, \citenamefont {Lonie},\ and\ \citenamefont
  {Titov}(1996)}]{Priest:1996A}%
  \BibitemOpen
  \bibfield  {author} {\bibinfo {author} {\bibfnamefont {E.~R.}\ \bibnamefont
  {Priest}}, \bibinfo {author} {\bibfnamefont {D.~P.}\ \bibnamefont {Lonie}}, \
  and\ \bibinfo {author} {\bibfnamefont {V.~S.}\ \bibnamefont {Titov}},\
  }\bibfield  {title} {\enquote {\bibinfo {title} {{Bifurcations of magnetic
  topology by the creation or annihilation of null points}},}\ }\href {\doibase
  10.1017/S0022377800019449} {\bibfield  {journal} {\bibinfo  {journal}
  {JPlPh}\ }\textbf {\bibinfo {volume} {56}},\ \bibinfo {pages} {507} (\bibinfo
  {year} {1996})}\BibitemShut {NoStop}%
\bibitem [{\citenamefont {Brown}\ and\ \citenamefont
  {Priest}(1999{\natexlab{a}})}]{DBrown:1999B}%
  \BibitemOpen
  \bibfield  {author} {\bibinfo {author} {\bibfnamefont {D.~S.}\ \bibnamefont
  {Brown}}\ and\ \bibinfo {author} {\bibfnamefont {E.~R.}\ \bibnamefont
  {Priest}},\ }\bibfield  {title} {\enquote {\bibinfo {title} {{Topological
  bifurcations in three-dimensional magnetic fields}},}\ }\href {\doibase
  10.1098/rspa.1999.0484} {\bibfield  {journal} {\bibinfo  {journal} {RSPSA}\
  }\textbf {\bibinfo {volume} {455}},\ \bibinfo {pages} {3931--3951} (\bibinfo
  {year} {1999}{\natexlab{a}})}\BibitemShut {NoStop}%
\bibitem [{\citenamefont {Brown}\ and\ \citenamefont
  {Priest}(2001)}]{DBrown:2001}%
  \BibitemOpen
  \bibfield  {author} {\bibinfo {author} {\bibfnamefont {D.~S.}\ \bibnamefont
  {Brown}}\ and\ \bibinfo {author} {\bibfnamefont {E.~R.}\ \bibnamefont
  {Priest}},\ }\bibfield  {title} {\enquote {\bibinfo {title} {{The topological
  behaviour of 3D null points in the Sun's corona}},}\ }\href {\doibase
  10.1051/0004-6361:20010016} {\bibfield  {journal} {\bibinfo  {journal}
  {Astron. Astrophys.}\ }\textbf {\bibinfo {volume} {367}},\ \bibinfo {pages}
  {339--346} (\bibinfo {year} {2001})}\BibitemShut {NoStop}%
\bibitem [{\citenamefont {Deimling}(1985)}]{deimling:1985}%
  \BibitemOpen
  \bibfield  {author} {\bibinfo {author} {\bibfnamefont {K.}~\bibnamefont
  {Deimling}},\ }\href@noop {} {\emph {\bibinfo {title} {{Nonlinear Functional
  Analysis}}}}\ (\bibinfo  {publisher} {{Springer-Verlag, New York}},\ \bibinfo
  {year} {1985})\BibitemShut {NoStop}%
\bibitem [{\citenamefont {Newcomb}(1958)}]{newcomb:1958}%
  \BibitemOpen
  \bibfield  {author} {\bibinfo {author} {\bibfnamefont {W.~A.}\ \bibnamefont
  {Newcomb}},\ }\bibfield  {title} {\enquote {\bibinfo {title} {{Motion of
  magnetic lines of force}},}\ }\href {\doibase 10.1016/0003-4916(58)90024-1}
  {\bibfield  {journal} {\bibinfo  {journal} {AnPhy}\ }\textbf {\bibinfo
  {volume} {3}},\ \bibinfo {pages} {347--385} (\bibinfo {year}
  {1958})}\BibitemShut {NoStop}%
\bibitem [{\citenamefont {Stern}(1966)}]{stern:1966}%
  \BibitemOpen
  \bibfield  {author} {\bibinfo {author} {\bibfnamefont {D.~P.}\ \bibnamefont
  {Stern}},\ }\bibfield  {title} {\enquote {\bibinfo {title} {{The Motion of
  Magnetic Field Lines}},}\ }\href {\doibase 10.1007/BF00222592} {\bibfield
  {journal} {\bibinfo  {journal} {Space Sci. Rev.}\ }\textbf {\bibinfo {volume}
  {6}},\ \bibinfo {pages} {147--173} (\bibinfo {year} {1966})}\BibitemShut
  {NoStop}%
\bibitem [{\citenamefont {Vasyliunas}(1972)}]{vasyliunas:1972}%
  \BibitemOpen
  \bibfield  {author} {\bibinfo {author} {\bibfnamefont {V.~M.}\ \bibnamefont
  {Vasyliunas}},\ }\bibfield  {title} {\enquote {\bibinfo {title}
  {{Nonuniqueness of magnetic field line motion}},}\ }\href {\doibase
  10.1029/JA077i031p06271} {\bibfield  {journal} {\bibinfo  {journal} {J.\
  Geophys.\ Res.}\ }\textbf {\bibinfo {volume} {77}},\ \bibinfo {pages} {6271}
  (\bibinfo {year} {1972})}\BibitemShut {NoStop}%
\bibitem [{\citenamefont {Greene}(1992)}]{greene:1992}%
  \BibitemOpen
  \bibfield  {author} {\bibinfo {author} {\bibfnamefont {J.~M.}\ \bibnamefont
  {Greene}},\ }\bibfield  {title} {\enquote {\bibinfo {title} {{Locating
  Three-Dimensional Roots by a Bisection Method}},}\ }\href {\doibase
  10.1016/0021-9991(92)90137-N} {\bibfield  {journal} {\bibinfo  {journal} {J.\
  Comput.\ Phys.}\ }\textbf {\bibinfo {volume} {98}},\ \bibinfo {pages} {194}
  (\bibinfo {year} {1992})}\BibitemShut {NoStop}%
\bibitem [{\citenamefont {Haynes}\ and\ \citenamefont
  {Parnell}(2007)}]{haynes:2007:trilinear}%
  \BibitemOpen
  \bibfield  {author} {\bibinfo {author} {\bibfnamefont {A.~L.}\ \bibnamefont
  {Haynes}}\ and\ \bibinfo {author} {\bibfnamefont {C.~E.}\ \bibnamefont
  {Parnell}},\ }\bibfield  {title} {\enquote {\bibinfo {title} {{A trilinear
  method for finding null points in a three-dimensional vector space}},}\
  }\href {\doibase 10.1063/1.2756751} {\bibfield  {journal} {\bibinfo
  {journal} {Phys.\ Plasmas}\ }\textbf {\bibinfo {volume} {14}},\ \bibinfo
  {pages} {082107} (\bibinfo {year} {2007})}\BibitemShut {NoStop}%
\bibitem [{\citenamefont {{Cullum}}(1971)}]{cullum:1971}%
  \BibitemOpen
  \bibfield  {author} {\bibinfo {author} {\bibfnamefont {J.}~\bibnamefont
  {{Cullum}}},\ }\bibfield  {title} {\enquote {\bibinfo {title} {{Numerical
  Differentiation and Regularization}},}\ }\href {\doibase 10.1137/0708026}
  {\bibfield  {journal} {\bibinfo  {journal} {SIAM Journal on Numerical
  Analysis}\ }\textbf {\bibinfo {volume} {8}},\ \bibinfo {pages} {254--265}
  (\bibinfo {year} {1971})}\BibitemShut {NoStop}%
\bibitem [{\citenamefont {Chartrand}(2011)}]{chartrand:2011}%
  \BibitemOpen
  \bibfield  {author} {\bibinfo {author} {\bibfnamefont {R.}~\bibnamefont
  {Chartrand}},\ }\bibfield  {title} {\enquote {\bibinfo {title} {Numerical
  differentiation of noisy, nonsmooth data},}\ }\href {\doibase
  10.5402/2011/164564} {\bibfield  {journal} {\bibinfo  {journal} {ISRN Applied
  Mathematics}\ }\textbf {\bibinfo {volume} {2011}},\ \bibinfo {eid} {164564}
  (\bibinfo {year} {2011}),\ 10.5402/2011/164564}\BibitemShut {NoStop}%
\bibitem [{\citenamefont {Edmondson}\ \emph {et~al.}(2010)\citenamefont
  {Edmondson}, \citenamefont {Antiochos}, \citenamefont {DeVore}, \citenamefont
  {Lynch},\ and\ \citenamefont {Zurbuchen}}]{edmondson:2010a}%
  \BibitemOpen
  \bibfield  {author} {\bibinfo {author} {\bibfnamefont {J.~K.}\ \bibnamefont
  {Edmondson}}, \bibinfo {author} {\bibfnamefont {S.~K.}\ \bibnamefont
  {Antiochos}}, \bibinfo {author} {\bibfnamefont {C.~R.}\ \bibnamefont
  {DeVore}}, \bibinfo {author} {\bibfnamefont {B.~J.}\ \bibnamefont {Lynch}}, \
  and\ \bibinfo {author} {\bibfnamefont {T.~H.}\ \bibnamefont {Zurbuchen}},\
  }\bibfield  {title} {\enquote {\bibinfo {title} {{Interchange Reconnection
  and Coronal Hole Dynamics}},}\ }\href {\doibase 10.1088/0004-637X/714/1/517}
  {\bibfield  {journal} {\bibinfo  {journal} {Astrophys. J.}\ }\textbf
  {\bibinfo {volume} {714}},\ \bibinfo {pages} {517--531} (\bibinfo {year}
  {2010})}\BibitemShut {NoStop}%
\bibitem [{\citenamefont {Shen}, \citenamefont {Lin},\ and\ \citenamefont
  {Murphy}(2011)}]{shen:2011}%
  \BibitemOpen
  \bibfield  {author} {\bibinfo {author} {\bibfnamefont {C.}~\bibnamefont
  {Shen}}, \bibinfo {author} {\bibfnamefont {J.}~\bibnamefont {Lin}}, \ and\
  \bibinfo {author} {\bibfnamefont {N.~A.}\ \bibnamefont {Murphy}},\ }\bibfield
   {title} {\enquote {\bibinfo {title} {{Numerical Experiments on Fine
  Structure within Reconnecting Current Sheets in Solar Flares}},}\ }\href
  {\doibase 10.1088/0004-637X/737/1/14} {\bibfield  {journal} {\bibinfo
  {journal} {Astrophys. J.}\ }\textbf {\bibinfo {volume} {737}},\ \bibinfo
  {pages} {14} (\bibinfo {year} {2011})}\BibitemShut {NoStop}%
\bibitem [{\citenamefont {Brown}\ and\ \citenamefont
  {Priest}(1999{\natexlab{b}})}]{DBrown:1999A}%
  \BibitemOpen
  \bibfield  {author} {\bibinfo {author} {\bibfnamefont {D.~S.}\ \bibnamefont
  {Brown}}\ and\ \bibinfo {author} {\bibfnamefont {E.~R.}\ \bibnamefont
  {Priest}},\ }\bibfield  {title} {\enquote {\bibinfo {title} {{The Topological
  Behaviour of Stable Magnetic Separators}},}\ }\href {\doibase
  10.1023/A:1005221503925} {\bibfield  {journal} {\bibinfo  {journal} {Sol.
  Phys.}\ }\textbf {\bibinfo {volume} {190}},\ \bibinfo {pages} {25--33}
  (\bibinfo {year} {1999}{\natexlab{b}})}\BibitemShut {NoStop}%
\bibitem [{\citenamefont {Vasyliunas}(1975)}]{vasyliunas:1975}%
  \BibitemOpen
  \bibfield  {author} {\bibinfo {author} {\bibfnamefont {V.~M.}\ \bibnamefont
  {Vasyliunas}},\ }\bibfield  {title} {\enquote {\bibinfo {title} {{Theoretical
  models of magnetic field line merging. I}},}\ }\href {\doibase
  10.1029/RG013i001p00303} {\bibfield  {journal} {\bibinfo  {journal} {RvGSP}\
  }\textbf {\bibinfo {volume} {13}},\ \bibinfo {pages} {303--336} (\bibinfo
  {year} {1975})}\BibitemShut {NoStop}%
\bibitem [{Note1()}]{Note1}%
  \BibitemOpen
  \bibinfo {note} {Null points are also known as neutral points, fixed points,
  stationary points, equilibrium points, critical points, singular points, and
  singularities. Separators are also known as saddle connectors and
  separation/attachment lines. Separatrix surfaces are also known as fans and
  separation surfaces.}\BibitemShut {Stop}%
\end{thebibliography}
\end{document}